\documentclass{article}
\usepackage{graphicx}
\usepackage{subfig}
\usepackage{floatrow}
\usepackage{verbatim} 
\usepackage{amsmath}
\usepackage{caption}
\usepackage{sidecap}
\usepackage{hyperref}
\usepackage[utf8]{inputenc}
\usepackage[margin=1in]{geometry}
\usepackage{authblk}
\usepackage{enumitem}

\title{From homogeneous to heterogeneous network alignment via colored graphlets}
\date{}
\begin{document}
\author[1]{Shawn Gu}
\author[1]{John Johnson}
\author[1,2]{Fazle E. Faisal}
\author[1,2,*]{Tijana Milenkovi\'{c}}
\affil[1]{Department of Computer Science and Engineering, University of Notre Dame, Notre Dame, IN, 46556, USA}
\affil[2]{Eck Institute for Global Health and Interdisciplinary Center for Network Science and Applications (iCeNSA), University of Notre Dame, Notre Dame, IN, 46556, USA}

\affil[*]{To whom correspondence should be addressed (email: tmilenko@nd.edu)}
\maketitle

\begin{abstract}
Network alignment (NA) compares networks with the goal of finding a node mapping that uncovers highly similar (conserved) network regions. Existing NA methods are homogeneous, i.e., they can deal only with networks containing nodes and edges of one type.  Due to increasing amounts of heterogeneous network data with nodes or edges of different types, we extend three recent state-of-the-art homogeneous NA methods, WAVE, MAGNA++, and SANA, to allow for heterogeneous NA for the first time. We introduce several algorithmic novelties. Namely, these existing methods compute homogeneous graphlet-based node similarities and then find high-scoring alignments with respect to these similarities, while simultaneously maximizing the amount of conserved edges. Instead, we extend homogeneous graphlets to their heterogeneous counterparts, which we then use to develop a new measure of heterogeneous node similarity. Also, we extend $S^3$, a state-of-the-art measure of edge conservation for homogeneous NA, to its heterogeneous counterpart. Then, we find high-scoring alignments with respect to our heterogeneous node similarity and edge conservation measures. In evaluations on synthetic and real-world biological networks, our proposed heterogeneous NA methods lead to higher-quality alignments and better robustness to noise in the data than their homogeneous counterparts. The software and data from this work is available upon request.
\end{abstract}

\section*{Introduction}

Due to advancements of biotechnologies for data collection, increasing amounts of biological network data are becoming available \cite{BioGRID, bamford2004cosmic, de2009aging, hulovatyy2014revealing}. A prominent type of biological networks is protein-protein interaction (PPI) networks. Aligning PPI networks of different species continues to be important \cite{sharan2006modeling, faisal2015post, emmert2016fifty, elmsallati2016global, guzzi2017survey}. This is because network alignment (NA) aims to uncover similar network regions by finding a node mapping between compared PPI networks. Then, analogous to genomic sequence alignment, NA can be used to transfer functional knowledge across species between their conserved PPI network (rather than sequence) regions. This is needed because functions of many proteins remain unknown even for well-studied species. Protein function prediction via NA-based across-species transfer can help close this gap.

NA methods typically consist of two main algorithmic components. First, the similarity between pairs of nodes from different networks is computed with respect to some measure of node conservation (NC). Second, an alignment strategy (AS) quickly identifies alignments that maximize total NC over all aligned nodes and the amount of conserved edges (i.e., edge conservation, EC). That is, intuitively, a good alignment should both map similar nodes to each other and preserve many edges.

Different types of NA methods exist. First, NA can be categorized as local (LNA) or global (GNA). LNA aims to find optimally conserved network regions, which typically results in the aligned regions being small \cite{berg2004local, berg2006cross, flannick2006graemlin, kelley2004pathblast, koyuturk2006pairwise, liang2006netalign, sharan2005conserved, ciriello2012alignnemo, mina2014improving}. On the other hand, GNA aims to find an overall node mapping between compared networks, which often results in the aligned network regions being large but suboptimally conserved \cite{faisal2015global, flannick2008automatic, klau2009new, kuchaiev2011integrative, kuchaiev2010topological, liao2009isorankn, milenkovic2010optimal, narayanan2011link, neyshabur2013netal, patro2012global, singh2007pairwise, singh2008global, zaslavskiy2009global}. Both LNA and GNA have  (dis)advantages \cite{meng2016local, meng2016igloo}. Since most of the recent work has dealt with GNA \cite{guzzi2017survey}, we also focus on GNA, but our work can be generalized to LNA as well.

Second, NA can be categorized as pairwise (PNA) or multiple (MNA). PNA is designed to find similar regions between exactly two networks, while MNA can align more than two networks. Because MNA is more computationally complex than PNA \cite{vijayanmilenkovic2017}, and because current PNA methods are also more accurate than current MNA methods \cite{vijayan2017pairwise}, we focus on PNA, but our work can be generalized to MNA as well.

Third, NA can be divided into two categories based on the type of its AS. One AS type is seed-and-extend, where first two highly similar  nodes (with respect to some NC measure) are aligned, i.e., seeded. Then, the seed's neighboring nodes (or simply neighbors) that are similar are aligned, the seed's neighbor's neighbors that are similar are aligned, and so on. This step of extending around the seed and exploring the seed's neighbors is intended to improve both NC and EC of the resulting alignment. The extension step continues until all nodes in the smaller of the two compared networks are aligned (formally, until a one-to-one node mapping between the two networks is produced). WAVE \cite{suncrawfordtangmilenkovic2015} is a state-of-the-art seed-and-extend AS, which was shown to work the best under a graphlet-based NC measure \cite{milenkovic2008,kuchaiev2010topological} (see below) and a score called ``weighted EC'', which is high if the nodes of the conserved edges (see below) are also similar with respect to the NC measure. The other AS type is a search algorithm. Here, instead of aligning node by node as with seed-and-extend ASs, entire alignments are explored and the one that scores the highest based on some objective function is returned. A typical objective function optimizes some measure of NC, EC, or a combination of the two. MAGNA++ \cite{MAGNAPP} and SANA \cite{SANA} are two state-of-the-art search algorithm-based ASs. MAGNA++ uses a genetic algorithm as its search strategy and it works the best under the objective function that optimizes the graphlet-based NC measure \cite{milenkovic2008,kuchaiev2010topological} and the $S^3$ EC measure \cite{MAGNAPP}. SANA uses simulated annealing as its search strategy, and it was evaluated under several objective functions that optimize EC, including $S^3$. In our study, we add to the EC (i.e., $S^3$) part of SANA's objective function the same graphlet-based NC measure that WAVE and MAGNA++ also optimize, in order to compare as fairly as possible the three NA methods and their heterogeneous counterparts.   

\begin{figure}[h]
    \centering
    \subfloat[]{\includegraphics[width=0.99\textwidth]{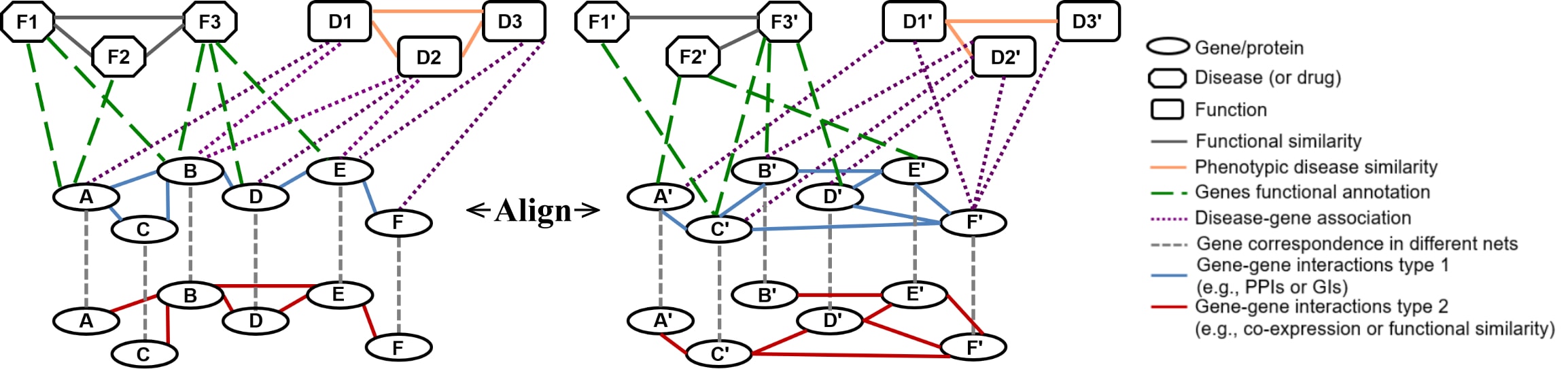}}\hfill
 \caption{\label{fig:hna-example}Illustration of two heterogeneous networks, each containing different node as well as edge types (or colors). In a given network, different node shapes represent different node types, and different line styles represent different edge types. If we do not consider the ovals with red edges (the bottom portion of the network), then we have a heterogeneous network with different node types, and thus implicitly different edge types. If we only consider the ovals with blue or red edges, then we have  a heterogeneous network with different edge types but a single node type (also called multimodal networks with two edge modes). The goal of HetNA as we define it is to find a node mapping between heterogeneous networks that contain different node types, different edge types, or both.}
    
\end{figure}

All existing NA methods are homogeneous (HomNA). That is, they deal with networks containing nodes and edges of one type. However, a network can have nodes or edges of more than one type (or color). For example, different  biological entities, such as proteins, phenotypes, or drugs, can be modeled as nodes, and different types of interactions, such as protein-protein, phenotype-phenotype, drug-drug, protein-phenotype, protein-drug, or phenotype-drug associations can be modeled as edges. Analyzing such heterogeneous multi-node- or multi-edge-type network data can lead to deeper insights into cellular functioning compared to homogeneous network analyses \cite{gligorijevic2015methods}. Therefore, there is a need for being able to perform heterogeneous NA (HetNA). Intuitively, HetNA aims to find a node mapping between heterogeneous networks (Figure \ref{fig:hna-example}). In this study, we propose the first ever approach for HetNA. 

While an existing method called AlignPI \cite{wu2009} was claimed to align heterogeneous networks, it actually did not perform HetNA as we define it in this study. Namely, AlignPI was simply used to align two networks of different types to one other (specifically, the human PPI network to the disease-disease association network). However, each of the two considered networks is homogeneous, and thus the networks were aligned in the homogeneous fashion. Another relevant existing method is Fuse \cite{FUSE}, which works via data integration. As such, it might appear that Fuse deals with data of different types, i.e., heterogeneous networks. However, it does not. Namely, Fuse aligns homogeneous PPI networks of different species, where the data integration step refers to using information from all of the homogeneous networks to calculate similarities between their nodes. Then, an alignment is still produced in the homogeneous fashion. The remaining relevant existing method is multimodal network alignment \cite{nassargleich2017}, which does deal with a special case of the HetNA problem. Namely, it aligns multimodal networks, which are a special case of heterogeneous networks as we define them. A multimodal (also called multiplex) network contains edges of different  types (or modes) between the same set of nodes. That is, it contains only a single node type (Figure \ref{fig:hna-example}). However, in our study, we define a heterogeneous network as a network that can contain different node types or different edge types (or both), and thus, our definition of HetNA is more broad than that of multimodal network alignment. Importantly, since the multimodal network alignment approach was not  published as of completion of our evaluation (i.e., it was available only on arXiv), the code implementing it was not available at the time. So, we were unable to consider this approach in our study.

\subsection*{Our contributions}

As already noted, current HomNA methods aim to find alignments with high homogeneous NC (HomNC) and  homogeneous EC (HomEC). So, to generalize HomNA to HetNA, we generalize HomNC to heterogeneous NC (HetNC) and HomEC to heterogeneous EC (HetEC). We describe these modifications intuitively below and formally in Methods.
\vspace{0.2cm}

\noindent\textbf{From homogeneous to heterogeneous NC.}  First, we introduce relevant concepts in the homogeneous context. Intuitively, two nodes from different homogeneous networks are topologically similar if their extended neighborhoods are similar. This idea can be quantified with homogeneous graphlets (small -- typically up to 5-node -- connected subgraphs), which have been been extensively studied in homogeneous network analysis \cite{milenkovic2008, yaverouglu2015proper, suncrawfordtangmilenkovic2015, hulovatyy2014revealing, solava2012graphlet, faisal2014dynamic, wang2014identification, singh2014graphlet}. For each node, for each graphlet, one counts how many times the given node touches each node symmetry group, or node orbit, in the given graphlet (e.g., in a 3-node path, the nodes at the end of the path are symmetric to each other and are thus in the same orbit, but they are distinct from the node in the middle, which is thus in a separate orbit). These counts over all graphlets summarize the extended network neighborhood of the node into its \textit{graphlet degree vector (GDV)}. Then, to compute topological similarity between two nodes, their GDVs are compared. 

\begin{figure}[h]
    \centering
    \subfloat[]{\includegraphics[width=0.3\textwidth]{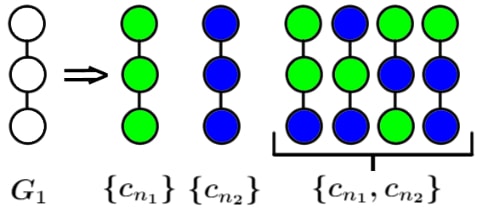}}
    \subfloat[]{\includegraphics[width=0.25\textwidth]{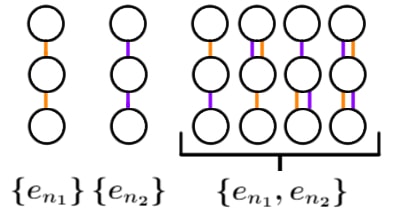}}
    
    \caption{\label{fig:col-graphlets-example}Illustration of (a) node-colored and (b) edge-colored graphlets. (a) With the exhaustive approach for enumerating all possible heterogeneous graphlets corresponding to homogeneous graphlet $G_1$, i.e., a 3-node path, given two colors, there would be six heterogeneous graphlets, each accounting for both which colors are present in the graphlet and which node position has which color. On the other hand, with our approach, there are three possible colored graphlets, denoted by $\{c_{n_1}\}$, $\{c_{n_2}\}$, and $\{c_{n_1},c_{n_2}\}$, each accounting only for which colors are present in the graphlet, ignoring the node-specific color information. Consequently, with our approach, the last four graphlets on the right of the arrow, which all have the same two colors present in them, are treated as the same heterogeneous graphlet. We design our approach in this way primarily to reduce the time complexity of counting heterogeneous graphlets in a network (but consequently, we also reduce the space complexity compared to the exhaustive approach). Namely, with our approach, the computational time complexity of searching for a given colored graphlet in a heterogeneous network remains the same as that of searching for its homogeneous equivalent. This is because the former involves: 1) counting in the heterogeneous network all graphlets, independent of their colors (which is the same as counting homogeneous graphlets in the network), and 2) for each of the homogeneous graphlets found in the network, simply determining which node colors appear in it and thus which node-colored graphlet the non-colored graphlet corresponds to. Step 1 is the time consuming part of the node-colored graphlet counting process, unlike step 2, which is trivial (can be done in constant time). (b) We develop a similar approach for edge-colored graphlets.}
\end{figure}

Second, when we have a heterogeneous (node- or edge-colored) network, we modify the above notion of topological similarity between nodes; now, two nodes from different networks are topologically similar if they are of the same color and if their extended neighborhoods are of similar color and network structure. To quantify this, we extend homogeneous graphlets into heterogeneous (or colored) graphlets, as follows. Given a heterogeneous network containing $n$ nodes and $c$ different node (or edge) colors, an exhaustive extension would track both which combinations of node (or edge) colors exist in a given graphlet as well as at which node (or edge) positions in the graphlet the colors occur. With such an approach, the computational complexity of the problem, namely both the enumeration of all possible heterogeneous graphlet types on up to $n$ nodes (the space complexity) and counting of the heterogeneous graphlets in a network (the time complexity), would increase exponentially with the number of colors   \cite{vacic2010graphlet}. Instead, we propose a more computationally efficient node-colored (or edge-colored) graphlet approach: we only track which combinations of node (or edge) colors exist in a given graphlet but not at which node (or edge) positions in the graphlet the colors occur  (Figure \ref{fig:col-graphlets-example}). Consequently, with our approach: 1) the number of possible colored graphlets and thus the computational space complexity is lower compared to the exhaustive approach, and 2) most importantly, the computational time complexity of counting colored graphlets in a heterogeneous network is the same as that of counting original graphlets in a homogeneous network, unlike with the exhaustive approach  (Figure \ref{fig:col-graphlets-example}). 
Given node- or edge-colored graphlets, analogous to the GDV of a node in a homogeneous network, we summarize the extended neighborhood of a node in a heterogeneous network with its \textit{node-colored GDV (NCGDV)} or \textit{edge-colored GDV (ECGDV)}. Then, we compute topological similarity between two nodes from heterogeneous networks by comparing the nodes' NCGDVs, ECGDVs, or both. Formal definitions of node-colored and edge-colored graphlets, as well as NCGDVs and ECGDVs, can be found in Methods.

Note that in our evaluation, we consider networks that contain  only different node types. As such, our considered data contain different edge types only implicitly, because edges between nodes of different types will by definition be of different types themselves. So, in our evaluation, we need to consider only node-colored graphlets and NCGDVs, but not edge-colored graphlets or ECGDVs. Yet, we propose, define, and provide software implementation for edge-colored graphlets and ECGDVs as well, because these can be used alone for alignment of multimodal networks or combined with node-colored graphlets and NCGDVs for alignment of heterogeneous networks such as those in Figure \ref{fig:hna-example}. 

The software implementing node-colored and edge-colored graphlet counting is available upon request. We also provide an intuitive graphical user interface (GUI) for easy use by domain scientists.

\vspace{0.2cm}

\noindent\textbf{From homogeneous to heterogeneous EC.}  In HomNA, $S^3$ is a state-of-the-art EC measure \cite{MAGNAPP, SANA}. To explain $S^3$, first, we need to define a conserved edge. Intuitively, given two nodes in one network, and given their aligned counterparts in another network, the alignment is said to conserve an edge (i.e., form a conserved edge) if the two nodes are connected in the first network and the aligned counterparts are connected in the other network. Otherwise, if only the two nodes in the first network are connected or only their aligned counterparts in the other network are connected, but not both, the alignment is said to not conserve an edge (i.e., form a non-conserved edge). Formal definitions of conserved and non-conserved edges can be found in Methods. Then, $S^3$ is defined the ratio of the number of conserved edges to the number of both conserved and non-conserved edges. Intuitively, $S^3$ rewards an alignment whenever it aligns an edge in one network to an edge in the other network and penalizes it whenever it aligns an edge in one network to a non-edge in the other network (or vice versa).   

We extend $S^3$ into a new measure of heterogeneous EC. In particular, we redefine what a conserved edge means, by accounting for colors of its aligned end nodes. Specifically, given a conserved edge consisting of nodes $u$ and $v$ in one network, and the corresponding aligned nodes $u'$ and $v'$, respectively, in the other network, if both $u$ and $u'$ have the same color and $v$ and $v'$ have the same color, then the edge is fully conserved. Instead, if either $u$ and $u'$ have the same color or $v$ and $v'$ have the same color, but not both, then the edge is partially conserved, i.e., its contribution to the heterogeneous $S^3$ score is penalized. If neither $u$ and $u'$ have the same color nor $v$ and $v'$ have the same color, then the edge is even less conserved than in the previous case, i.e., its contribution to the heterogeneous $S^3$ score is penalized even more. Finally, if the edge is non-conserved, we treat it the same as in the homogeneous case. In this way, our new heterogeneous $S^3$ measure favors both conserving edges and conserving edges whose aligned end nodes match in color.

\begin{figure}[h]
{\caption{\label{fig:EC-example}Illustration of HomEC and HetEC for an alignment between networks $G$ and $H$. Arrows represent one possible alignment (mapping) between the networks, i.e., their nodes. Note that this node mapping is not the best alignment possible with respect to HomEC, but we use it to illustrate the concepts involved. For a detailed illustration of conserved edges, non-conserved edges, and $S^3$, see Section Introduction--From homogeneous to heterogeneous EC.}

}
{\includegraphics[width=5cm]{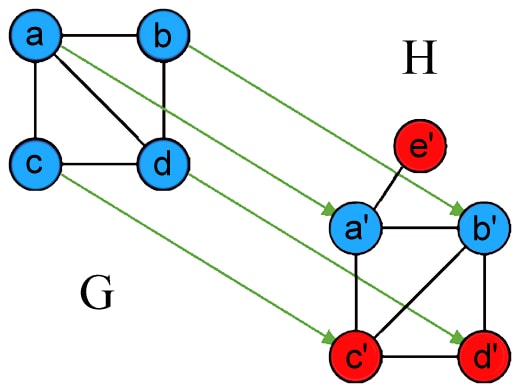}}
\end{figure}

Here we give a concrete example of these concepts for the alignment in Figure \ref{fig:EC-example}. In the homogeneous case (i.e., if all nodes were of the same color), there exist four conserved edges: the one formed by $(a, a)$ and $(a', a')$ -- because $a$ is aligned to $a'$, $b$ is aligned to $b'$, and an edge exists both between $a$ and $b$ as well as between $a'$ and $b'$; the one formed by $(a, c)$ and $(a', c')$; the one formed by $(c, d)$ and $(c', d')$; and the one formed by $(b, d)$ and $(b', d')$. On the other hand, $(a, d)$ and $(a', d')$ form a non-conserved edge, because while $a$ is aligned to $a'$ and $d$ is aligned to $d'$, there is an edge between $a$ and $d$ but not between $a'$ and $d'$. For a similar reason, $(b, c)$ and $(b', c')$ form another non-conserved edge. So, given the existence of four conserved edges and two non-conserved edges, homogeneous $S^3$ is $\frac{\text{\# conserved edges}}{(\text{\# conserved edges} + \text{\# non-conserved edges})} = 4 / (4 + 2) = 0.67$. In the heterogeneous case, for an edge to be conserved, the homogeneous condition is still required. However, we also account for colors of the aligned end nodes of a conserved edge and penalize for color mismatches. Specifically, $(a, b)$ and $(a', b')$ are counted as a fully conserved edge (with conservation weight of 1), because in addition to the fact that this edge is conserved in the homogeneous case, $a$ has the same color as $a'$, and $b$ has the same color as $b'$. $(a, c)$ and $(a', c')$ are counted as a less conserved edge (with conservation weight of $\frac{2}{3}$), because while $a$ and $a'$ have the same color, $c$ and $c'$ do not. Similarly, $(b, d)$ and $(b', d')$ form a partly conserved edge with conservation weight of  $\frac{2}{3}$. $(c, d)$ and $(c', d')$ are counted as an even less conserved edge (with conservation weight of $\frac{1}{3}$) because neither $c$ and $c'$ nor $d$ and $d'$ have the same color. Finally, $(a, d)$ and $(a', d')$ form a non-conserved edge, just as in the homogeneous case. Given the total edge conservation of $1 + \frac{2}{3} + \frac{2}{3} + \frac{1}{3} = \frac{8}{3}$ and two non-conserved edges (the same ones as in the homogeneous case), heterogeneous $S^3$ uses the same formula as $S^3$ and is $\frac{8}{3} / (\frac{8}{3} + 2) = 0.57$.

\vspace{0.2cm}

\noindent\textbf{From homogeneous to heterogeneous NA.} We modify existing HomNA methods WAVE, MAGNA++, and SANA to perform HetNA by optimizing our new HetNC and HetEC measures (instead of their original HomNC and HomEC measures) with these methods' ASs. We choose WAVE and MAGNA++ because they rose to the top in the study by Meng \emph{et al.}, 2016 \cite{meng2016local}, which is a recent comprehensive evaluation of 10 HomNA methods. Since then, SANA appeared and was promising. So, we include SANA into our study as well. We modify all three methods and evaluate their new heterogeneous versions as described below. Detailed descriptions of these methods and their heterogeneous modifications can be found in Methods.

\section*{Results}

First, we describe our evaluation framework, specifically data that we use, networks that we align, and parameters of the three considered NA methods. Second, we compare HomNA and HetNA. That is, we compare each of homogeneous WAVE, MAGNA++, and SANA to its heterogeneous counterpart. Recall that there currently exist no HetNA methods, and thus, we cannot compare heterogeneous WAVE, MAGNA++, or SANA to any other hetNA method except to each other. In more detail, we evaluate: 1) the effect of HetNC, i.e., whether using more node colors increases alignment quality (and especially whether using two or more colors, i.e., HetNA, is superior to using a single color, i.e., HomNA), 2) the effect of HetEC, i.e., whether using heterogeneous $S^3$ over homogeneous $S^3$ increases alignment quality, and 3) the effect of the alignment method, i.e., which of our three new HetNA methods performs the best with respect to accuracy and running time. 

\subsection*{Evaluation}

We perform three evaluation tests corresponding to three sets of networks: 1) synthetic networks with up to four artificially imposed node colors, 2) homogeneous human PPI networks that have up to four node colors imposed according to proteins' involvement in a combination of aging, cancer, and Alzheimer disease (AD), and 3) heterogeneous human protein-GO networks, where the two node colors correspond to proteins and their Gene Ontology (GO) terms, and edges exist between proteins, between proteins and GO terms, and between GO terms. Note that while we evaluate WAVE and SANA in all three tests, due to MAGNA++'s computational complexity, we evaluate MAGNA++ only in the first test on the smaller synthetic networks but not in the remaining two tests on the larger PPI or protein-GO networks. We align each of the above networks to its noisy versions. Details are as follows.

\vspace{0.2cm}

\noindent\textbf{Synthetic networks.} We form synthetic networks using two random graph generators, namely: 1) geometric random graphs \cite{penrose2003random} (GEO) and 2) scale-free networks \cite{barabasi1999emergence} (SF). The two models have distinct network topologies \cite{milenkovic2008graphcrunch}, which enables us to test the robustness of our results to the choice of random graph model. We form five random network instances per model and average results over them to account for the stochastic nature of the models. We set all model network instances to the same size of 1,000 nodes and 6,000 edges. Since the existing random graph generators are not designed to produce heterogeneous networks, we simply randomly assign each node a color out of $k$ possible colors, where there are approximately $1000/k$ nodes of each color. We vary $k$ from one to four. That is, for each synthetic network, we form heterogeneous versions with one, two, three, and four colors. 

\vspace{0.2cm}

\noindent\textbf{Human PPI networks.} We obtain the human PPI network data from BioGRID \cite{BioGRID}. We consider two types of PPIs: only affinity capture coupled to mass spectrometry (APMS) and only two-hybrid (Y2H). Sizes of the resulting networks are shown in Table \ref{tab:ppi-sizes}.

\begin{table}[h]
\begin{center}
\begin{tabular}{ c c c }
 \hline
 Network & \# of nodes & \# of edges  \\
 \hline
 APMS & 11,450 & 92,257 \\ 
 Y2H & 10,317 & 41,925 \\  
 \hline

\end{tabular}
\caption{Number of nodes and edges in the two considered PPI networks.}\label{tab:ppi-sizes}
\end{center}
\end{table}

We impose node colors onto each PPI network based on the proteins' involvement in a combination of aging, cancer, and Alzheimer's disease (AD). We obtain a list of sequence-based (Seq) human aging-related genes from GenAge \cite{de2009aging} and a list of gene expression-based (Expr) human aging-related genes from the study by Berchtold et al., 2008 \cite{berchtold2008gene}. We obtain a list of genes related in cancer from COSMIC \cite{bamford2004cosmic}. We obtain a list of human genes related to AD from Simpson et al., 2011 \cite{simpson2011microarray}. 

We use these data to impose colors onto nodes in each of the two PPI networks (as well as their noisy counterparts; see below). For a given network, we use sequence-based aging- and cancer-related data to form four different colored versions of the network, as follows:

\setlist[itemize]{noitemsep, topsep=0pt}
\begin{itemize}

    \item In the 1-colored network, we treat all the nodes the same, meaning they have the same color.
    \item In the 2-colored network, we use the aging-related data to color nodes as ``aging-related". Otherwise, they are ``non-aging-related". This gives us 270 ``aging-related" and 10,047 ``non-aging-related" nodes.
    \item In the 3-colored network, we use aging- and cancer-related data. If a node is present in the aging-related data, we color it ``aging-related". If a node is absent there but present in the cancer-related data, we color it as ``cancer only". If a node is absent from both, we color it as ``non-aging-related and non-cancer". In this way, we have 270 ``aging-related", 405 ``cancer only", and 9,642 "non-aging-related and non-cancer" nodes.
    \item In the 4-colored network, we use the same scheme as the 3-colored network, except if a node is present in both data sets, we color it as ``both aging-related and cancer". This gives us 203 ``aging-related", 405 ``cancer only", 67 ``both aging-related and cancer", and 9,642 ``non-aging-related and non-cancer" nodes.
    
\end{itemize}

To test the robustness of the choice of node color data above, we vary the underlying data. Now, for each of the two PPI network types, we use expression-based aging- and AD-related data to form four colored versions of the given network, as follows:

\begin{itemize}

    \item In the 1-colored network, we treat all the nodes the same, meaning they have the same color.
    \item In the 2-colored network, we use the aging-related data to color nodes as ``aging-related". Otherwise, they are ``non-aging-related". This gives us 2,889 ``aging-related" and 7,428 ``non-aging-related" nodes.
    \item In the 3-colored network, we use aging- and AD-related data. If a node is present in the aging-related data, we color it ``aging-related". If a node is absent there but present in the AD-related data, we color it as ``AD only". If a node is absent from both, we color it as ``non-aging-related and non-AD". In this way, we have 2,889 ``aging-related", 356 ``AD only", and 7,072 ``non-aging-related and non-AD" nodes.
    \item In the 4-colored network, we use the same scheme as the 3-colored network, except if a node is present in both data sets, we color it as ``both aging-related and AD". This gives us 2,232 ``aging-related", 356 ``AD only", 657 ``both aging-related and AD", and 7,072 ``non-aging-related and non-AD" nodes.
    
\end{itemize}

\vspace{0.2cm}

\noindent\textbf{Human protein-GO networks.}  A heterogeneous protein-GO network has two types of nodes: protein and GO term \cite{ashburner2000gene}, and three types of edges: 1) PPI, 2) protein-GO association, and 3) GO-GO semantic similarity. The PPI data are the same two types of PPI networks as before (APMS and Y2H), protein-GO associations are obtained from the Gene Ontology Consortium \cite{ashburner2000gene} based on experimental evidence codes, and GO-GO semantic similarities are computed as follows. We compute semantic similarity between all GOs that annotate at least one protein in the given considered PPI network. We use Lin method \cite{mazandu2013dago} to compute the semantic similarity. We form edges between GOs using semantic similarity threshold of 0.7, because the density of the resulting GO-GO network approximately matches the density of the corresponding PPI network.
Considering APMS PPIs only and Y2H PPIs only, we form two heterogeneous protein-GO networks for human, whose sizes are shown in Tables \ref{tab:prot-GO-nodes} and \ref{tab:prot-GO-edges}.

\begin{table}[h]
\begin{center}
\begin{tabular}{ c c c c }
 \hline
 Network & & Node type \\
 & \# of proteins & \# of GO terms & \# of all nodes combined \\
 \hline
 APMS & 11,450 & 5,558 & 17,008 \\ 
 Y2H & 10,317 & 5,554 & 15,871 \\  
 \hline

\end{tabular}
\caption{Number of nodes in the two considered heterogeneous protein-GO networks.}\label{tab:prot-GO-nodes}
\end{center}
\end{table}

\begin{table}[h]
\begin{center}
\begin{tabular}{ c c c c c }
 \hline
 Network & \multicolumn{4}{c}{Edge type} \\ 
 & \# of PPIs & \# of protein-GO associations & \# of GO-GO semantic similarities & \# of all edges combined \\
 \hline
 APMS & 92,257 & 24,854 & 48,731 & 165,842 \\ 
 Y2H & 41,925 & 24,473 & 48,873 & 115,271 \\  
 \hline

\end{tabular}
\caption{Number of edges in the two considered heterogeneous protein-GO networks.}\label{tab:prot-GO-edges}
\end{center}
\end{table}

\vspace{0.2cm}

\noindent\textbf{Creating noisy counterparts of a synthetic, PPI, or protein-GO network.}  Given an original network $G$, we construct its noisy counterparts as follows. Considering a noise level of $x\%$, we randomly choose $x\%$ of the edges  and remove them from the original network, and then we randomly choose the same number of node pairs that are disconnected in the original network and add edges between them. That is, we randomly rewire $x\%$ of the edges in the original network. Each noisy network has the same number of nodes and edges as the original network. For each considered original network, we use the following noise levels: 0\%, 10\%, 25\%, 50\%, 75\%, and 100\%. We construct multiple instances of noisy networks at each level to account for the randomness in edge rewiring; then, we average results (i.e., alignment quality) over the multiple runs. For WAVE and SANA, we use at least three instances. For MAGNA++, we only use one instance due to MAGNA++'s high computation complexity. 

\vspace{0.2cm}

\noindent\textbf{Measuring alignment quality.}  Since we align an original network to its noisy counterpart, we know the true node mapping between the aligned networks (of course, this mapping is hidden from each NA method when it is asked to produce an alignment). Therefore, we evaluate the quality of the given network by measuring its node correctness, which quantifies how well the alignment matches the true node mapping. Formally, node correctness is the percentage of node pairs from the given alignment that are present in the true node mapping. 

\subsection*{Comparison of HomNA and HetNA}

We need to define our considered evaluation scenarios. HomNA uses HomNC and HomEC, and we call this scenario HomNC-HomEC. For HetNA, if HetNC is used with HomEC, we call this scenario HetNC-HomEC; if HomNC is used with HetEC, we call this scenario HomNC-HetEC; and if HetNC is used with HetEC, we call this scenario HetNC-HetEC. 
Note that while MAGNA++ and SANA can optimize both NC and EC because they are search algorithms, WAVE only optimizes NC and it cannot directly optimize EC, because it is a seed-and-extend algorithm. Hence, while we can evaluate MAGNA++ and SANA in all four of the above scenarios, i.e., while for these two methods we can study  the effect on alignment quality of both HomNC versus HetNC and HomEC versus HetEC, for WAVE, we can only study the effect of HomNC versus HetNC.

First, we compare  HomNC-HomEC to HetNC-HomEC, to study the effect of HetNC alone on alignment quality, while still considering HomEC in both cases. Then, we compare HetNC-HomEC to HetNC-HetEC to study the effect of HetEC on alignment quality after we have already accounted for HetNC. We perform all of these comparisons comprehensively, using all considered methods on all considered data sets, as described in Methods. We also compare  HomNC-HomEC to HomNC-HetEC to additionally study the effect of HetEC on alignment quality without first accounting  for HetNC. Here, we perform only several case study comparisons out of all possible comparisons, due to the already comprehensive comparison experiments mentioned above.

\vspace{0.2cm}

\begin{figure}[h!]
    \centering
    \subfloat[]{\includegraphics[width=0.31\textwidth]{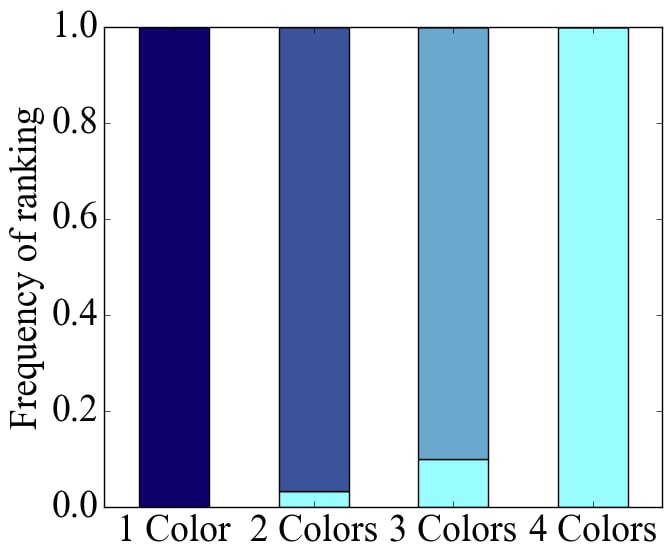}}\hfill
    \subfloat[]{\includegraphics[width=0.29\textwidth]{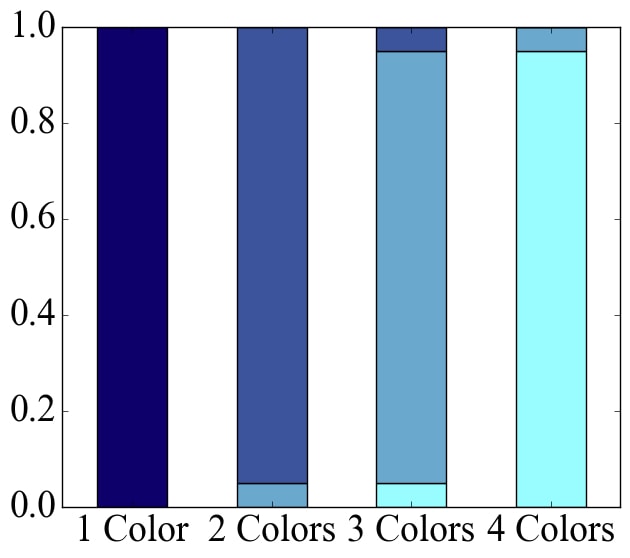}}\hfill
    \subfloat[]{\includegraphics[width=0.383\textwidth]{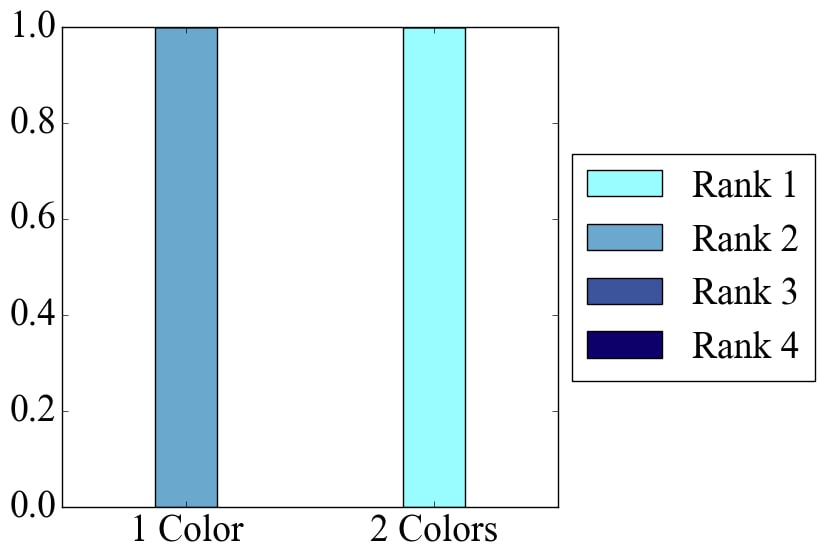}}\hfill
    \caption{\label{fig:col-sum}Summarized results regarding the effect of the \textbf{number of considered node colors} on alignment quality for (a) synthetic networks, (b) PPI networks, and (c) protein-GO networks. In panels (a) and (b), there are up to four considered node colors, while in panel (c), there are up to two considered node colors (see Section Evaluation for details). For each case (see below), we compare the different color levels (i.e., numbers of considered colors shown on \textit{x}-axes) and rank them from the best (rank 1) to the worst (rank 4 in panels a and b, and rank 2 in panel c). Then, we compute the percentage or frequency of all cases (see below) in which the given color level is ranked as the first (rank 1), second (rank 2), third (rank 3), or fourth (rank 4) best among all considered color levels. In panel (a), there are 3 methods (WAVE, MAGNA++, SANA) $\times$ 2 networks (geometric, scale-free) $\times$  5 noise levels (0\%, 10\%, 25\%, 50\%, 75\%) = 30 cases. In panel (b), there are 2 methods (WAVE, SANA) $\times$  4 networks (APMS-Expr, APMS-Seq, Y2H-Expr, Y2H-Seq) $\times$ 5 noise levels (0\%, 10\%, 25\%, 50\%, 75\%) = 40 cases. In panel (c), there are 2 methods (WAVE, SANA) $\times$ 2 networks (protein-GO-APMS, protein-GO-Y2H) $\times$  5 noise levels (0\%, 10\%, 25\%, 50\%, 75\%) = 20 cases.  Note that we analyzed an additional noise level (100\%), but we leave the corresponding results from this summary figure, because at this level all cases are expected to result in the same (random) alignments (Section Evaluation--Creating noisy counterparts of a synthetic, PPI, or protein-GO network). Instead, we show the results for the noise level of 100\% in the detailed figures (Figures \ref{fig:3-methods-synth}, \ref{fig:2-methods-ppi},  \ref{fig:2-methods-prot-go}). Also, note that in this figure, for each case, we choose the best between HetNC-HomEC and HetNC-HetEC.} 
\end{figure}

\begin{figure}[h!]
    \centering
    \subfloat[]{\includegraphics[width=0.29\textwidth]{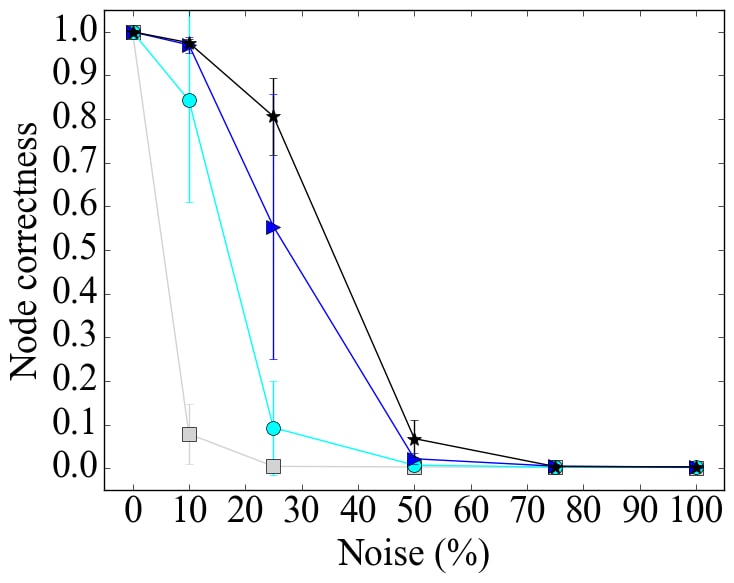}}\hfill
    \subfloat[]{\includegraphics[width=0.27\textwidth]{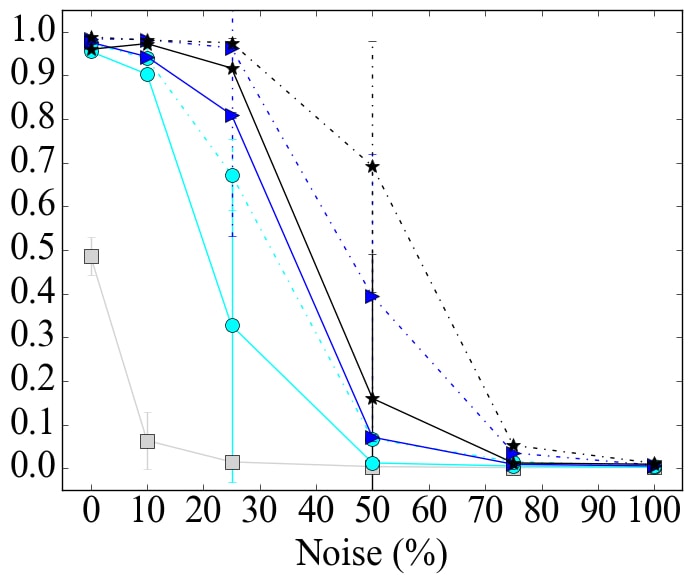}}\hfill
    \subfloat[]{\includegraphics[width=0.422\textwidth]{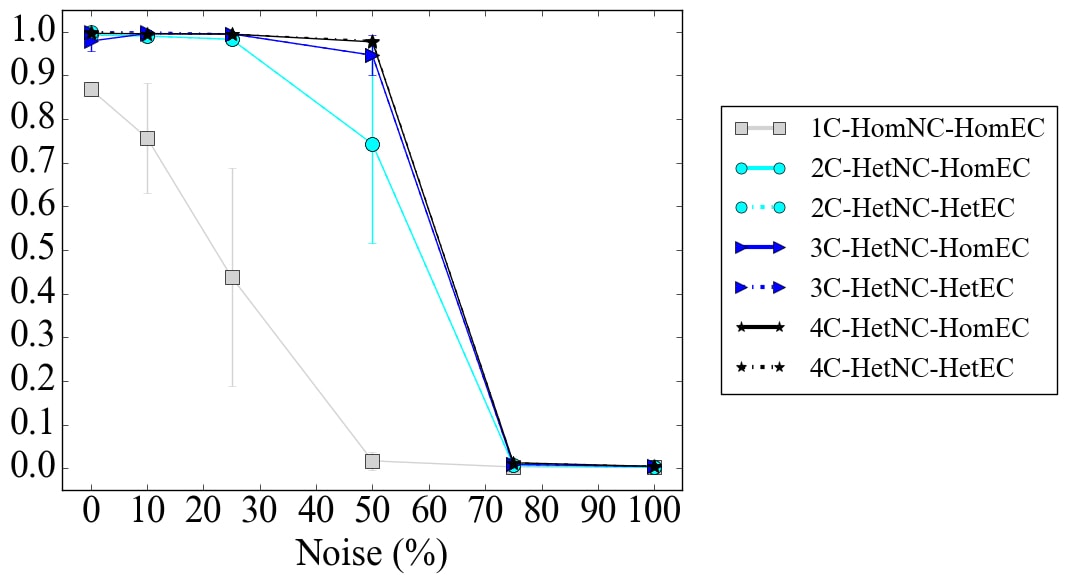}}\hfill
    \caption{\label{fig:3-methods-synth}Detailed alignment quality results regarding the effect of the \textbf{number of node colors} on alignment quality as a function of noise level for \textbf{synthetic, specifically geometric}, networks, using (a) WAVE, (b) MAGNA++, and (c) SANA. Gray squares, light blue circles, dark blue triangles, and black stars indicate the aligned networks containing one, two, three, and four node colors, respectively. For two or more node colors, solid lines represent using HetNC-HomEC, and dashed lines represent using HetNC-HetEC. Equivalent results for the remaining synthetic, specifically scale-free, networks are shown in Supplementary Figure S2.}
\end{figure}

\begin{figure}[h!]
    \centering
    \subfloat[]{\includegraphics[width=0.29\textwidth]{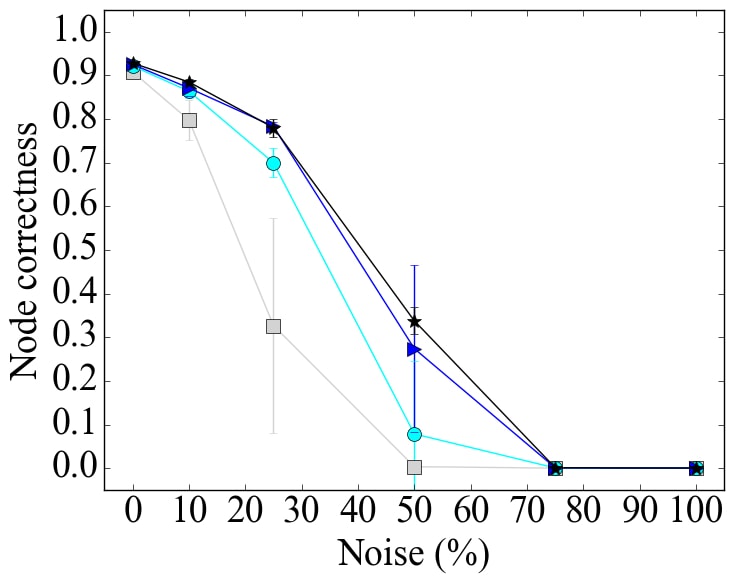}}
    \subfloat[]{\includegraphics[width=0.422\textwidth]{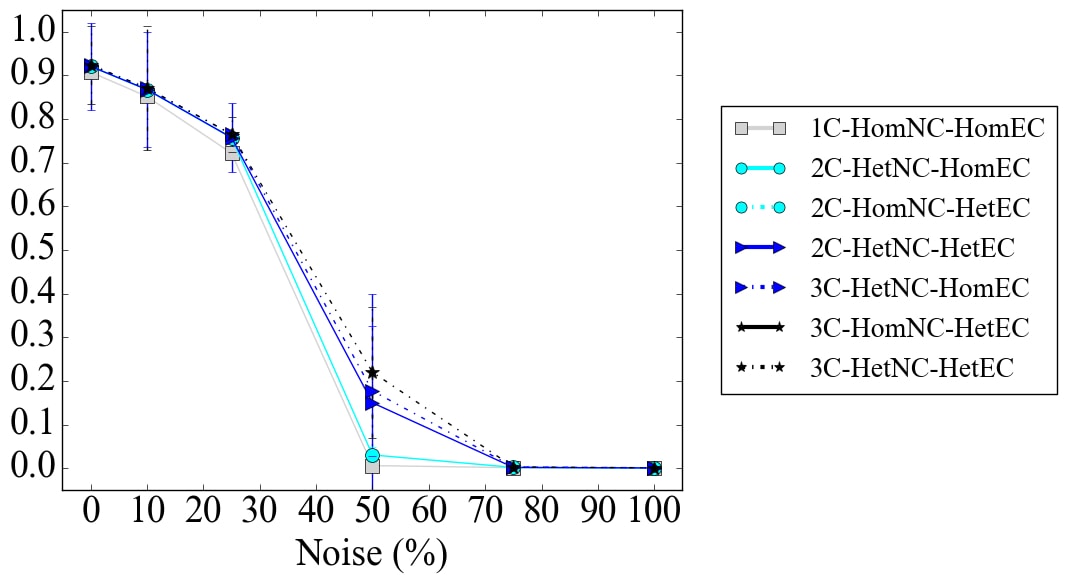}}
    \caption{\label{fig:2-methods-ppi}Detailed alignment quality results regarding the effect of the \textbf{number of node colors} on alignment quality as a function of noise level for \textbf{PPI, specifically APMS-Expr}, networks using (a) WAVE and (b) SANA. The figure can be interpreted in the same way as Figure \ref{fig:3-methods-synth}. Recall that for these larger networks, we have not run MAGNA++ due to its high computational complexity. Equivalent results for the remaining PPI, specifically APMS-Seq, Y2H-Expr, and Y2H-Seq, networks are shown in Supplementary Figures S4, S5, and S6.}
\end{figure}

\begin{figure}[h!]
    \centering
    \subfloat[]{\includegraphics[width=0.29\textwidth]{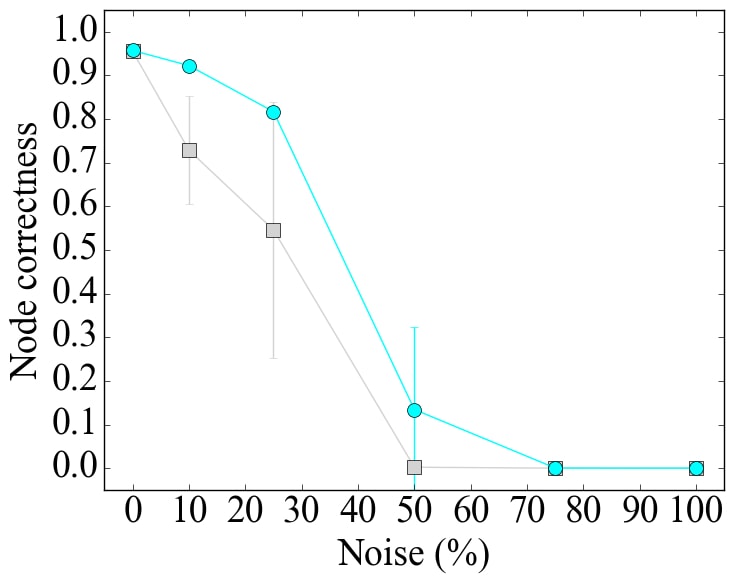}}
    \subfloat[]{\includegraphics[width=0.422\textwidth]{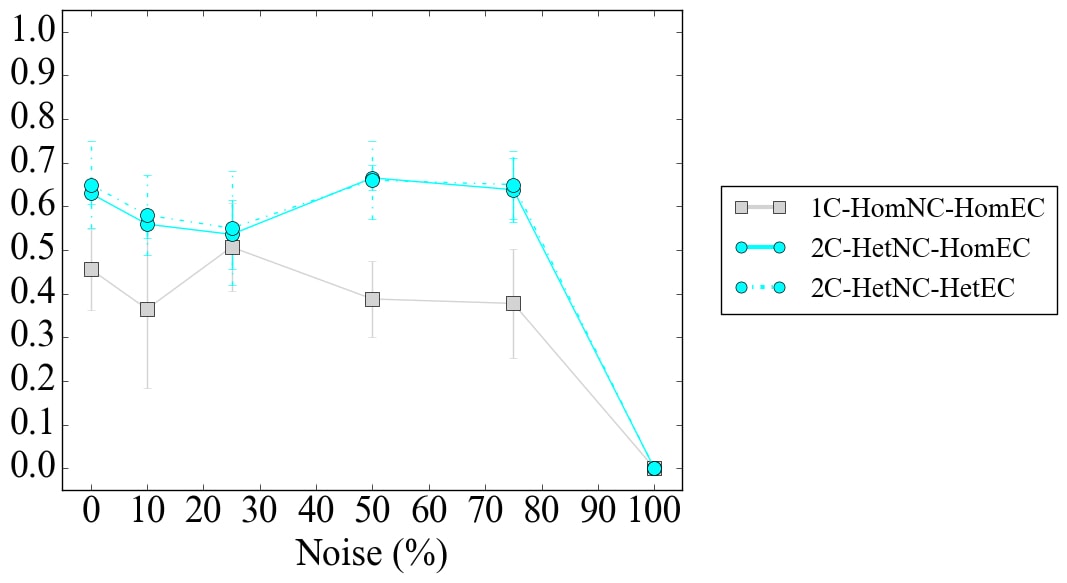}}
    \caption{\label{fig:2-methods-prot-go}Detailed alignment quality results regarding the effect of the \textbf{number of node colors} on alignment quality as a function of noise level for \textbf{protein-GO, specifically protein-GO-APMS}, networks using (a) WAVE and (b) SANA. The figure can be interpreted in the same way as Figure \ref{fig:3-methods-synth}. Recall that for these larger networks, we have not run MAGNA++ due to its high computational complexity. Equivalent results for the remaining protein-GO, specifically protein-GO-Y2H, networks are shown in Supplementary Figure S8.}
\end{figure}

\noindent\textbf{The effect of HetNC.} In terms of accuracy, we expect that for a given noise level, HetNA (i.e., HetNC-HomEC or HetNC-HetEC -- two or more node colors) should improve alignment quality over HomNA (i.e., HomNC-HomEC -- one node color). Also, we expect that the more colors are used, the better the alignment quality should be, since more information is used in the process of producing the alignment. In addition, we predict that using more colors will make the given method more robust to noise, meaning that we should see a slower decrease in alignment quality as noise increases, compared to using fewer colors. However, alignment quality should be low at the highest noise levels regardless of how many colors we use, since we are essentially aligning two networks with almost random topologies compared to each other. Indeed, we observe these exact trends (Figures \ref{fig:col-sum}, \ref{fig:3-methods-synth}, \ref{fig:2-methods-ppi}, \ref{fig:2-methods-prot-go}). Note that the few observed ties occur typically at the lower (0\% and 10\%) noise levels, which makes sense because in such cases network similarity can be captured reliably, meaning that all methods perform well.

In terms of time complexity, due to the way we count homogeneous as well as heterogeneous graphlets, time does not increase with more colors. Because of this, and because using more colors results in higher accuracy, we recommend using as many colors as needed. Note, however, that space complexity increases with the increase in the number of considered colors, because there are more possible graphlets; yet, the space complexity is practically feasible for a reasonable number of colors, such as four considered colors in our study (Section Methods--From homogeneous to heterogeneous NC). 


\vspace{0.2cm}

\begin{figure}[h!]
    \centering
    \subfloat[]{\includegraphics[width=0.31\textwidth]{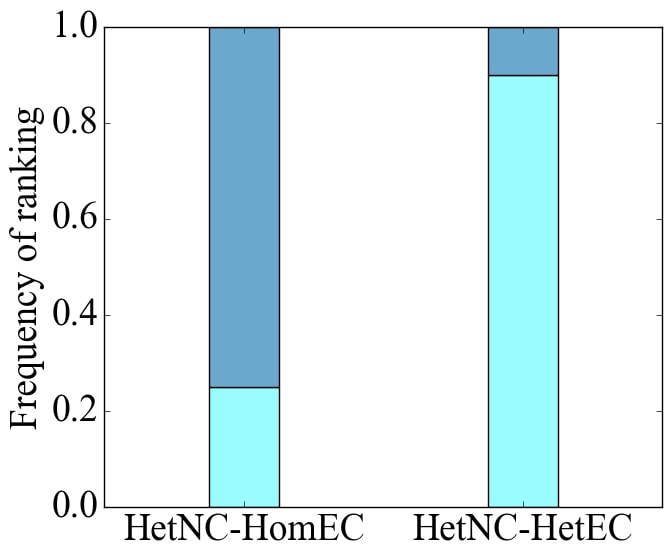}}\hfill
    \subfloat[]{\includegraphics[width=0.29\textwidth]{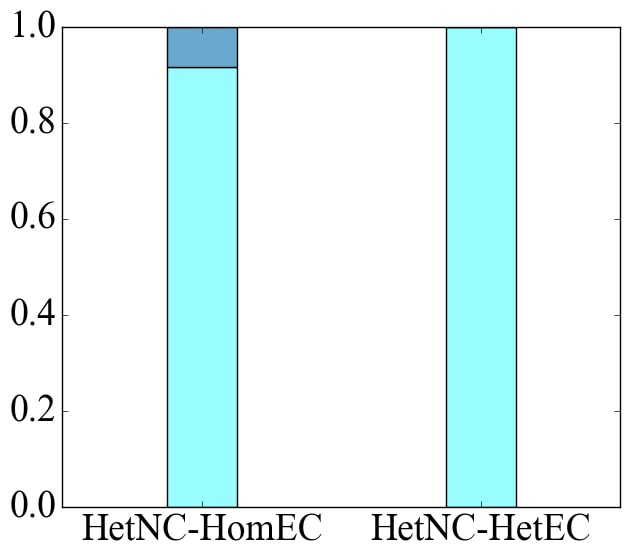}}\hfill
    \subfloat[]{\includegraphics[width=0.383\textwidth]{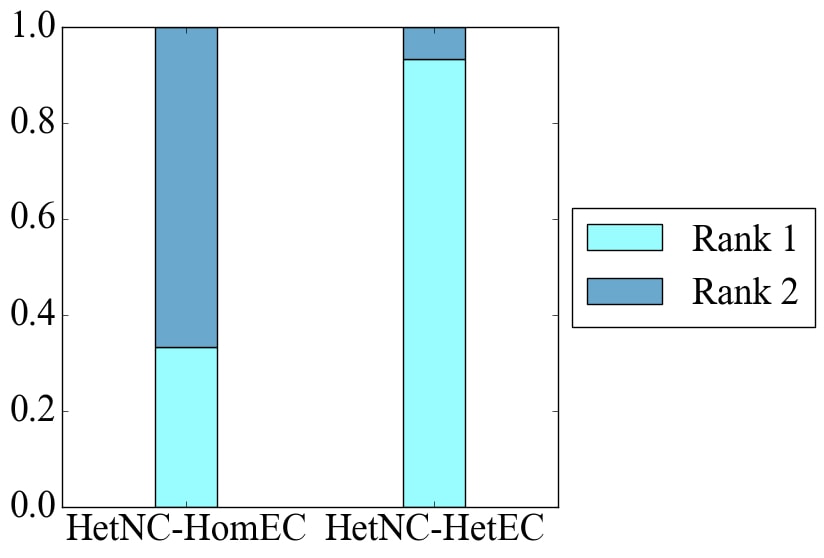}}\hfill
    \caption{\label{fig:het-sum-all} Summarized results regarding the effect of using \textbf{HetEC over HomEC} (both with HetNC) on alignment quality for (a) synthetic networks, (b) PPI networks, and (c) protein-GO networks. In all panels, there are two evaluation scenarios (HetNC-HomEC and HetNC-HetEC). For each case (see below), we compare the two considered evaluation scenarios and rank them from the best (rank 1) to the worst (rank 2). Then, we compute the percentage or frequency of all cases (see below) in which the given scenario is ranked as the first (rank 1) and second (rank 2) best among the considered scenarios. In panel (a), there are 2 methods (MAGNA++, SANA) $\times$ 2 networks (geometric, scale-free) $\times$ 5 noise levels (0, 10, 25, 50, 75) $\times$ 3 colors (1 color does not have a HetEC counterpart) = 60 cases. In panel (b), there is 1 method (SANA) $\times$ 4 networks (APMS-Expr, APMS-Seq, Y2H-Expr, Y2H-Seq) $\times$ 5 noise levels (as before) $\times$ 3 colors (as before) = 60 cases. In panel (c), there is 1 method (SANA) $\times$ 2 networks (protein-GO-APMS, protein-GO-Y2H) $\times$ 5 noise levels (as before) $\times$ 1 color (maximum 2 colors, but 1 color does not have a HetEC counterpart) = 10 cases. Note that we analyzed an additional noise level (100\%), but we leave the corresponding results from this summary figure, because at this level all cases are expected to result in the same (random) alignments (Section Evaluation--Creating noise counterparts of a synthetic, PPI, or protein-GO network). Instead, we show the results for the noise level of 100\% in the detailed figures (Figures \ref{fig:3-methods-synth}, \ref{fig:2-methods-ppi}, \ref{fig:2-methods-prot-go}).}
\end{figure}

\noindent\textbf{The effect of HetEC.}
In terms of accuracy, we expect improvement of HetNC-HetEC over HetNC-HomEC, because while both HomEC and HetEC favor aligning nodes that conserve edges, unlike HomEC, HetEC also favors aligning nodes whose colors match. Indeed, this is generally what we observe (Figure  \ref{fig:het-sum-all}). 

However, we see some ties between HetNC-HomEC and HetNC-HetEC. Also, while for MAGNA++ HetNC-HetEC noticeably improves alignment quality over HetNC-HomEC, for SANA, improvements of HetNC-HetEC over HetNC-HomEC are usually small (Figures \ref{fig:3-methods-synth}, \ref{fig:2-methods-ppi}, \ref{fig:2-methods-prot-go}). (WAVE does not explicitly optimize EC, so we are unable to compare HomEC versus HetEC for WAVE). This could be due to SANA's algorithm: it explores millions of alignments a second, and thus, it seems to already find high-scoring ones with just HetNC, without the need for HetEC.

For these reasons,  we consider the HomNC-HetEC scenario, to properly gauge the true potential of HetEC in the task of HetNA, without any ``bias" of also already using HetNC. Here, we analyze only two cases as a proof-of-concept of the effect of HetEC while still considering HomNC. Specifically, the two cases are MAGNA++ on geometric networks and SANA on APMS-Expr networks.

\begin{figure}[h!]
    \centering
    \subfloat[]{\includegraphics[width=0.29\textwidth]{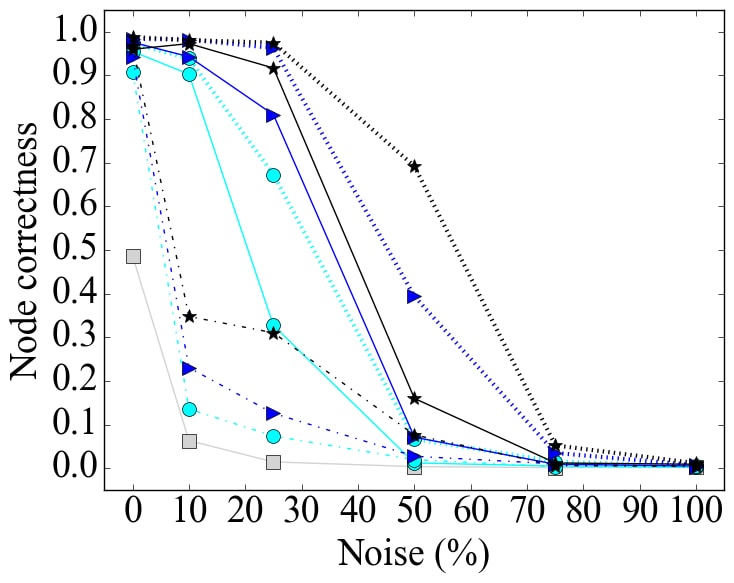}}
    \subfloat[]{\includegraphics[width=0.422\textwidth]{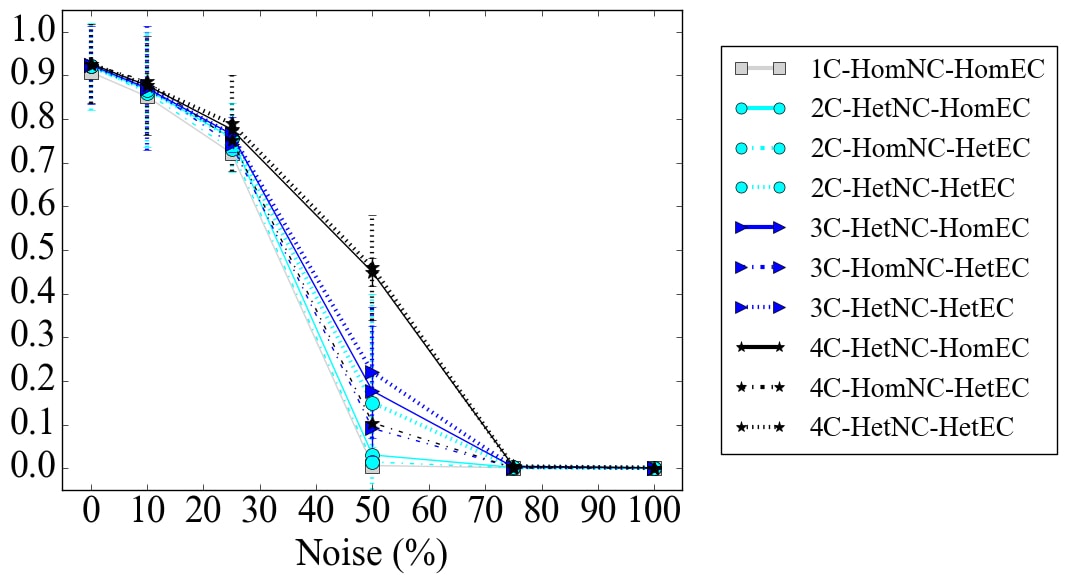}}
    \caption{Detailed alignment quality results regarding the effect of \textbf{HomNC-HetEC} compared to HomNC-HomEC, HetNC-HomEC, and HetNC-HetEC on alignment quality for the two considered case study evaluation tests: (a) geometric networks using MAGNA++ and (b) APMS-Expr networks using SANA. The figure can be interpreted in the same way as Fig. \ref{fig:3-methods-synth}, except that now solid lines represent HetNC-HomEC, short-long dotted lines represent HomNC-HetEC, and finely dotted lines represent HetNC-HetEC.}\label{fig:case-study}
\end{figure}

For these two cases, we evaluate all of HomNC-HomEC, HetNC-HomEC, HomNC-HetEC, and HetNC-HetEC scenarios (Figure \ref{fig:case-study}). First, for a given scenario, for a given noise level, we ask whether using more colors yields higher alignment quality, as expected. Indeed, this is what we observe. Second, for both MAGNA++ and SANA, HomNC-HetEC improves over HomNC-HomEC (i.e., over HomNA), though for SANA improvements are again small. However, using HetNC alone (HetNC-HomEC) improves alignment quality more than using HetEC alone (HomNC-HetEC). This might not be surprising, because HetNC favors aligning nodes of the same color that also have similar extended neighborhoods, while HetEC does not account for this extended neighborhood. As expected, HetNC-HetEC yields the best alignment quality of all four cases for all colors and all noise levels, except the highest (75\% and 100\%), as expected. For MAGNA++ on geometric networks, the improvements of HetNC-HetEC over the next best scenario (HetNC-HomEC) are large, while for SANA on APMS-Expr networks, the improvements over the next best scenario (also HetNC-HomEC) are marginal.

In terms of time complexity, calculating heterogeneous $S^3$ (i.e., HetEC) has the same complexity as calculating homogeneous $S^3$ (i.e., HomEC), since counting the number of conserved and non-conserved edges in a heterogeneous network takes the same amount of time as in a homogeneous network. Specifically, checking if node colors match (Section Introduction--From homogeneous to heterogeneous EC) to determine how much conserved an edge is takes constant time. Because of this, and because using both HetNC and HetEC results in the highest accuracy, we recommend using both HetNC and HetEC (i.e., HetNC-HetEC scenario).

\begin{figure}[h!]
    \centering
    \subfloat[]{\includegraphics[width=0.31\textwidth]{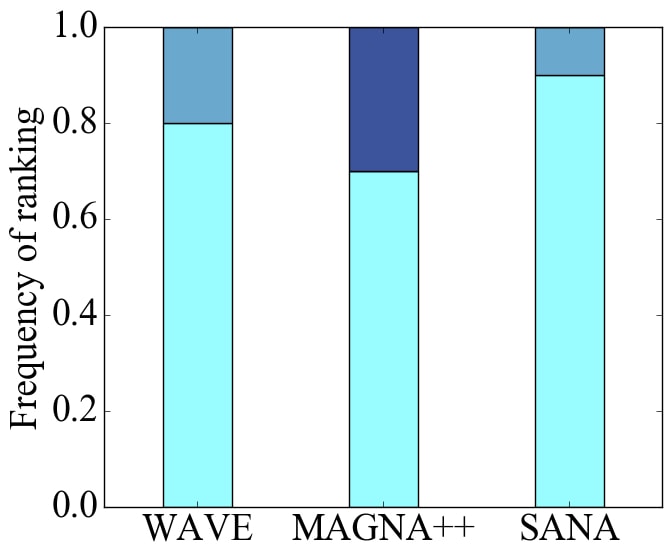}}\hfill
    \subfloat[]{\includegraphics[width=0.29\textwidth]{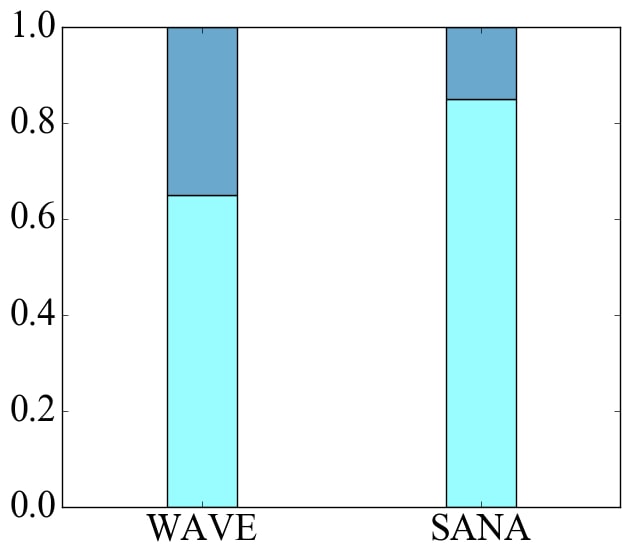}}\hfill
    \subfloat[]{\includegraphics[width=0.383\textwidth]{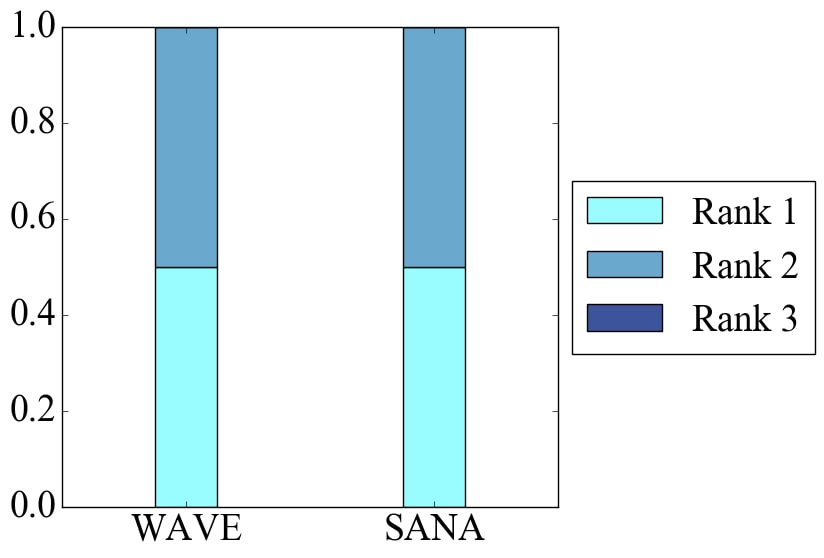}}\hfill
    \caption{\label{fig:method-sum}Summarized results regarding the effect of the \textbf{alignment method} on alignment quality for (a) synthetic networks, (b) PPI networks, and (c) protein-GO networks. In panel (a), there are three considered alignment methods (WAVE, MAGNA++, and SANA). In panels (b) and (c), there are two considered alignment methods (WAVE and SANA). For each case (see below), we compare the alignment methods and rank the different methods from best (rank 1) to worst (rank 3 in panel (a), and rank 2 in panels (b) and (c)). Then, we compute the percentage of all cases in which the given method is ranked as the first, second, or third best among all considered methods. In panel (a), there are 2 networks (geometric, scale-free) $\times$ 5 noise levels (0, 10, 25, 50, 75) = 10 cases. In panel (b), there are 4 networks (APMS-Expr, APMS-Seq, Y2H-Expr, Y2H-Seq) $\times$ 5 noise levels (as above) = 20 cases. In panel (c), there are 2 networks (protein-GO-APMS, protein-GO-Y2H) $\times$ 5 noise levels (as above) = 10 cases. Note that we analyzed an additional noise level (100\%), but we leave the corresponding results from this summary figure, because at this level all cases are expected to result in the same (random) alignments (Section Evaluation-Creating noise counterparts of a synthetic, PPI, or protein-GO network). Instead, we show the results for the noise level of 100\% in the detailed figures (Figures \ref{fig:3-methods-synth}, \ref{fig:2-methods-ppi}, \ref{fig:2-methods-prot-go}). Also, note that in this figure, we give each method the best case advantage. That is, we show results for the best of HetNC-HomEC and HetNC-HetEC, and also only for the maximum node color level (four colors in panels (a) and (b), and two colors in panel (c)). We do the latter because of all color levels, it is the maximum color level at which the given method performs the best, for each method. Nonetheless, the results remain qualitatively the same if we account for all considered colored levels. }
\end{figure}

\begin{figure}[h!]
    \subfloat[]{\includegraphics[width=0.33\textwidth]{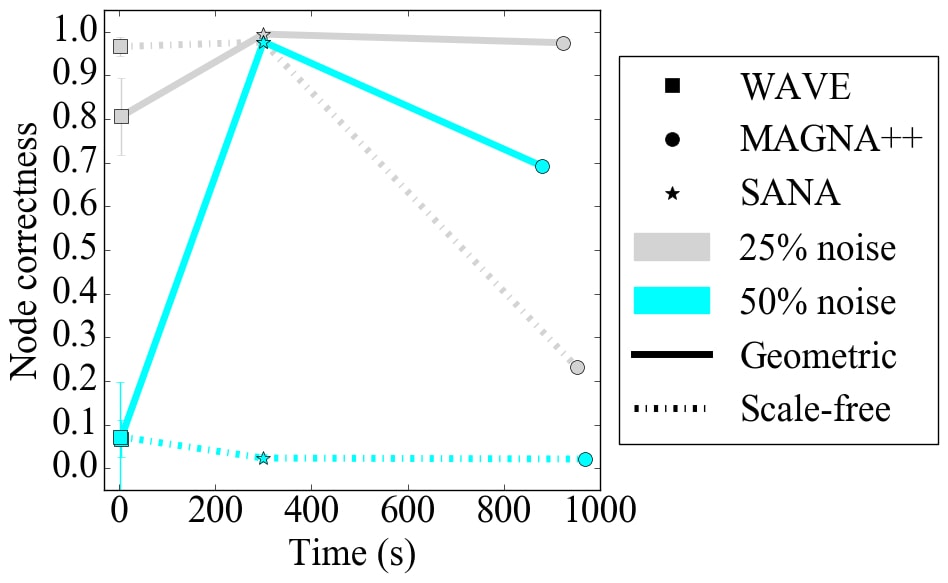}}
    \subfloat[]{\includegraphics[width=0.33\textwidth]{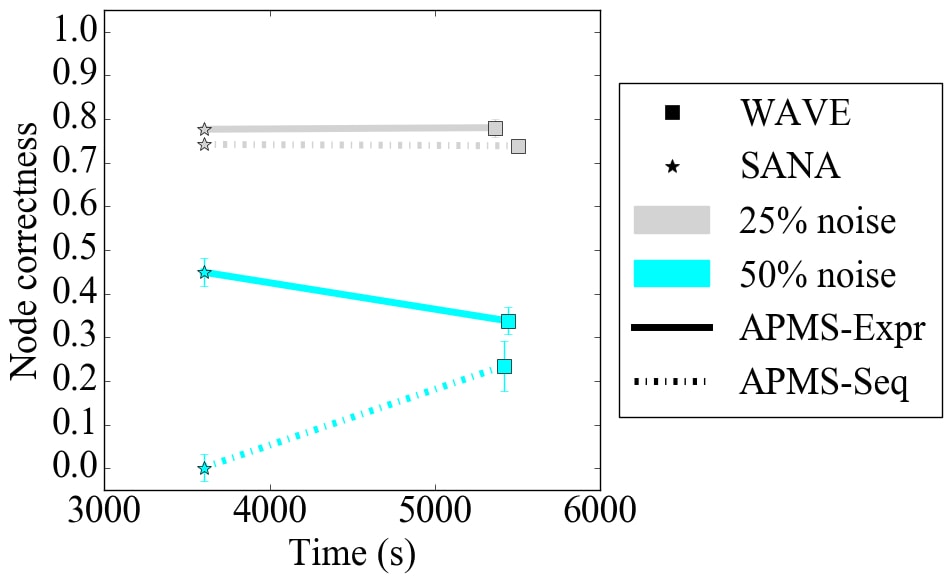}}
    \subfloat[]{\includegraphics[width=0.33\textwidth]{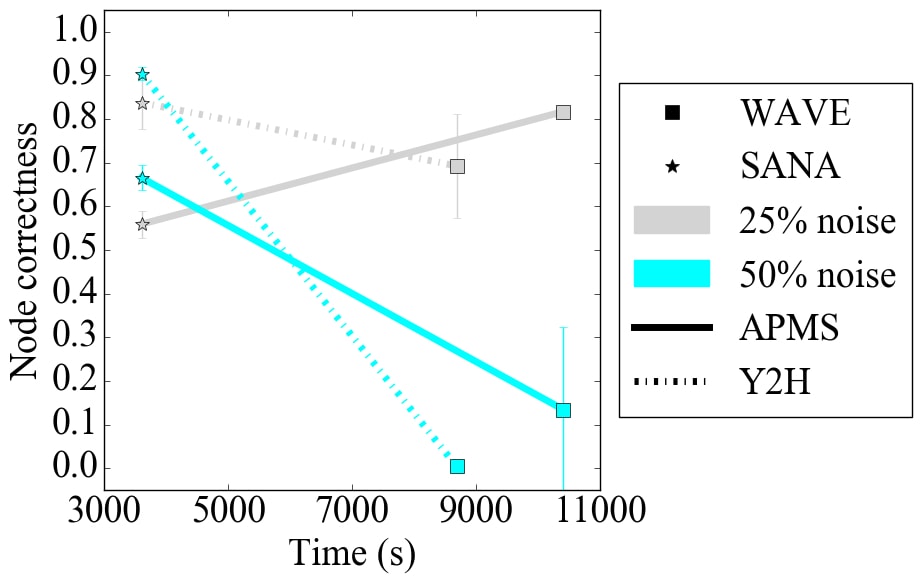}}
    \caption{\label{fig:time-meth-sum}Summarized results comparing the \textbf{running times verus accuracy} of different methods for 25\% and 50\% noise on (a) synthetic, specifically geometric and scale-free, (b) PPI, specifically APMS-Expr and APMS-Seq, and (c) protein-GO, specifically APMS and Y2H, networks. The \textit{x}-axis is the running time of the given method on the given network data at the given noise level, and the \textit{y}-axis is the alignment quality score. Here we use different shapes to represent the different methods, different colors to represent the different noise levels, and solid or broken lines to represent the different network data. Lines are drawn between the different methods for the same noise level and network data, for easier comparison of the different methods. Detailed running time results for all other noise levels and network data are shown in Supplementary Figures S9-S16.}
\end{figure}

\vspace{0.2cm}
\noindent\textbf{The effect of alignment method.}  In terms of accuracy, regardless of noise level, WAVE and SANA generally outperform MAGNA++ (Figure \ref{fig:method-sum}). WAVE and SANA have somewhat comparable performance (Fig. \ref{fig:method-sum}), in the following sense. For synthetic networks, the two are tied in 70\% of all evaluation tests, WAVE is superior to SANA in 10\% of the tests,  and SANA is superior to WAVE in 20\% of the tests. For PPI networks, the two are tied in 50\% of all evaluation tests, WAVE is superior to SANA in 15\% of the tests, and SANA is superior to WAVE in 35\% of the tests. For protein-GO networks, the two are tied in 0\% of all evaluation tests, WAVE is superior to SANA in 50\% of the tests, and SANA is superior to WAVE in 50\% of the tests. Whenever WAVE is superior to SANA, it is typically for lower noise levels (up to 25\%) (Figure \ref{fig:time-meth-sum}). Whenever SANA is superior to WAVE, it is typically for higher noise levels (above 25\%) (Figure \ref{fig:time-meth-sum}). These trends for lower versus higher noise levels could be due WAVE's algorithm. At lower noise levels, the networks being aligned are still very similar to each other, so if two nodes are topologically similar, then it is likely that they should be aligned to each other. In this situation, WAVE would start with a good seed and thus be likely to produce a good alignment. At higher noise levels, the networks being aligned are dissimilar. So, two nodes may be topologically similar only because of the random rewiring of edges, but still be (erroneously) mapped to each other. In this situation, WAVE would start with a poor seed and likely lead to a poor alignment. Since SANA is not a seed-and-extend method, it avoids this issue and performs well even at higher noise levels.

In terms of time complexity, MAGNA++ is the slowest of the three methods (Fig. \ref{fig:time-meth-sum}(a)), which is expected since it uses a genetic algorithm. Of WAVE and SANA, for synthetic networks, which happen to be the smallest of our considered networks, WAVE is faster than SANA (Figure \ref{fig:time-meth-sum}(a)). However, keep in mind that the execution time is a parameter in SANA. In that sense, it is possible to run SANA so that it is faster than any other method. However, in this case, SANA might not reach desired alignment quality. It might be possible to run SANA for as long as needed to always beat or at least tie WAVE in terms of alignment quality, but the amount of time would have to be determined empirically for every network pair being aligned. For PPI and protein-GO networks, which happen to be the largest of our considered networks, SANA is faster than WAVE (Figure \ref{fig:time-meth-sum}(b)-(c)).

\section*{Discussion}

We modify WAVE, MAGNA++, and SANA to align heterogeneous networks by extending the existing notions of NC and EC to their heterogeneous counterparts. Specifically, we extend homogeneous graphlets to their heterogeneous counterparts, and homogeneous $S^3$ to heterogeneous $S^3$. We evaluate our methods by aligning synthetic, PPI, and protein-GO networks to their noisy counterparts. We show that using more colors leads to better alignments, and that using both heterogeneous NC and heterogeneous EC is the preferred option where available. Also, we find that WAVE and SANA perform equally well at lower noise levels, though SANA does better at higher noise levels. 

There are many new directions in which this work could be taken. Faster heterogeneous graphlet counting methods could be developed by using combinatorial relationships between heterogeneous graphlets, akin to existing efficient methods for homogeneous graphlet counting \cite{hovcevar2014combinatorial, marcus2012rage, rahman2014graft, ahmed2015efficient}. Or, faster, more scalable methods for capturing the topology of a node in a heterogeneous network could be developed as an alternative to graphlets, such as those based on random walks \cite{grover2016node2vec, dong2017metapath2vec}. Also, our considered networks have up to four colors; aligning networks with more colors, as well as adding explicit (rather than just implicit, as in our study) edge colors, could show further improvements. Another direction is improving the AS of NA methods. For example, in WAVE, the choice of the first aligned (seed) node pair likely impacts the rest of the alignment. If there are many possibilities for this pair, can an algorithm discover the best one, independent of the noise level in the data? Furthermore, while NA has been extended from dealing with static networks to dealing with dynamic networks \cite{vijayan2017alignment, vijayan2017aligning}, the existing dynamic NA work currently only deals with homogeneous dynamic networks. Developing methods to align heterogeneous dynamic networks may yield improvements. In a similar vein, our current heterogeneous work deals with PNA, and so extending it into heterogeneous MNA may be of future interest.

\section*{Methods}

\subsection*{Calculating node similarities, i.e., NC}

Given the GDV for each node in a network, we form a matrix of GDVs over all nodes for each of the two networks being aligned. Then, we combine the two matrices row-wise and perform PCA on the large matrix of the networks' GDVs. We choose the first $r$ principal components, where $r$ is at least two and as small as possible such that the $r$ components account for at least 90\% of the variation in the data. Then, for every pair of nodes between the two networks, we calculate their cosine similarity based on the nodes' principal components and scale so the values are between $0$ and $1$.
\vspace{0.2cm}

\noindent\textbf{Method parameters.} WAVE does not have any parameters. We set MAGNA++'s parameters as follows: we use initial population size of 15,000 and 2,000 generations, which are the suggested values in the MAGNA++ documentation; we run MAGNA++ on 16 threads on all networks. We give equal weight to MAGNA++'s NC and EC measures, i.e., we set its \texttt{a} parameter to 0.5; using this value has been suggested by several studies \cite{MAGNAPP,meng2016local}. We set SANA's parameters as follows: we give equal weight to its NC and EC measures for fair comparability with MAGNA++, i.e., we set the following parameters: \texttt{s3} (corresponding to EC) to 1, \texttt{esim} (corresponding to NC) to 1, \texttt{simFile} to the name of the NC-based node similarity file, and \texttt{simFormat} to 1 (this tells SANA to read the similarity file such that each line has 3 columns: node1, node2, and the similarity between them). SANA also has a parameter for how long it should search for alignments. For synthetic networks, we run  SANA for the default 5 minutes (\texttt{t} 5). For PPI and protein-GO networks, we increase the \texttt{t} parameter to 60 minutes (\texttt{t} 60), since these networks are larger and thus SANA needs more time to find a good alignment (which we have verified empirically in our evaluation). 

\subsection*{From homogeneous to heterogeneous NC} 

Here we formalize the notion of heterogeneous (colored) graphlets. For ease of explanation, first, we define node-colored graphlets. 
Given $k$ possible node colors from the set $C_n = \{c_{n_1}, c_{n_2}, ..., c_{n_k}\}$, $S = 2^{C_n}$ is the set of all possible combinations of colors from $C_n$. $S$ contains $\binom{k}{0}$ elements with no color (i.e. the empty set), $\binom{k}{1}$ elements with any one color, and in general $\binom{k}{i}$ elements with any $i$ colors. Therefore, $S$ contains $2^k$ elements. So $S \setminus \emptyset$ is the set of all possible color combinations from $C_n$ that excludes the empty set, which contains $2^k - 1$ elements. Let $b_n \in S \setminus \emptyset$. Given a homogeneous graphlet $G_i$, a set of colors $C_n$, and some $b_n$, define a node-colored graphlet $NCG_{i, b_n}$ to be the set of all distinct graphs that are isomorphic to $G_i$, such that for each graph, each node is colored with one of the colors from $b_n$, and also,  each color from $b_n$ has to be present in each such graph. Thus, given $k$ node colors, there are $2^k-1$ possible node-colored graphlets.

As an illustration, let us assume that a heterogeneous network has nodes with two possible colors: $c_{n_1}$ and $c_{n_2}$. These two node colors have 3 possible combinations: $\{c_{n_1}\}$, $\{c_{n_2}\}$, and $\{c_{n_1}, c_{n_2}\}$. As a result, for each homogeneous graphlet $G_i$, there are three possible node colored graphlets (Figure \ref{fig:col-graphlets-example}).  

This definition of node-colored graphlets is more space efficient than the exhaustive approach is: given a heterogeneous network containing $n$ nodes and $k$ different colors, with the exhaustive approach, both the number of possible colored graphlets (the space complexity) and the the time needed to count such graphlets in the network (the time complexity) increase exponentially with the number of colors.  With our approach, however, 1) the number of possible colored graphlets is much smaller (though still exponential in terms of the number of colors) compared to the exhaustive approach, and 2) the  time complexity of counting colored graphlets in a heterogeneous network is the same as that of counting original graphlets in a homogeneous network, unlike with the exhaustive approach.

Regarding the space complexity of our colored graphlet approach, as an illustration, for two colors, with the exhaustive definition, there would be six node-colored graphlets for homogeneous graphlet $G_1$, a 3-node path, while with our approach there are only three of them. For three colors, with the exhaustive definition, there would be 18 node-colored graphlets for $G_1$, while with our approach there are only seven of them. Although even with our approach, the number of node-colored graphlets increases drastically with the increase of $k$, but this is not a major concern because in practice we may expect a relatively small value of $k$. For example, one can study a heterogeneous network whose nodes are proteins, functions, diseases, and drugs with $k$ value of only four.

Just as an orbit (i.e., topological symmetry group) of a homogeneous graphlet \cite{milenkovic2008}, we define an orbit of a node-colored graphlet $NCG_{i,b_n}$ as the set of nodes that are ``symmetric" to each other in $NCG_{i,b_n}$; the symmetry  ignores node colors (Figure \ref{fig:col-graphlets-example}). For a homogeneous graphlet with $x$ orbits, each of its colored graphlets also has $x$ orbits. That is, given $k$ node colors, there are $73 \times (2^k - 1)$ orbits for 2-5-node node-colored graphlets (there are 73 orbits for homogeneous 2-5 node graphlets). Then, we define heterogeneous \textit{node-colored GDV (NCGDV)} by counting the number of node-colored graphlets that the given node ``touches" at each of the node-colored orbits. Analogous to the homogeneous case, to compare two nodes in heterogeneous networks, we compare their NCGDVs. 

Second, analogous to the definitions for node-colored graphlets, without going again through all the formalisms, we define edge-colored graphlets (Fig. \ref{fig:col-graphlets-example}), orbits in edge-colored graphlets, and \textit{edge-colored GDV (ECGDV)}. In practice, we may expect a relatively small number of edge colors (e.g., we can study a network whose nodes are genes/proteins and whose edges are PPIs, genetic interactions, gene co-expressions, and signaling interactions with only four edge colors).

Third, the above ideas can be combined to define truly heterogeneous graphlets that have different node and edge colors. For each node-colored graphlet, one can vary its edge colors. Alternatively, it is possible and computationally much simpler to concatenate NCGDVs and ECGDVs, which does not add any additional computational complexity compared to computing only NCGDVs or only ECGDVs.

\subsection*{From homogeneous to heterogeneous EC}

Let $u, v$ be two nodes in a network $G$, and $u', v'$ be two nodes in a network $H$. Let $f$ be a mapping (i.e., alignment) from the nodes of $G$ to the nodes of $H$ such that $f(u) = u'$ and $f(v)  = v'$ (another way to say this is that source node $u$ has image $u'$, and source node $v$ has image $v'$). That is, $u$ is aligned to $u'$, and $v$ is aligned to $v'$. Then, a conserved edge is formed by two edges from different networks such that each end node of one edge is aligned under $f$ to a unique end node of the other edge. On the other hand, a non-conserved edge is formed by an edge from one network and a pair of nodes from the other network that do not form an edge, such that each end node of the edge is aligned under $f$ to a unique node of the non-edge. Then, homogeneous $S^3$ of an alignment is defined as the ratio of conserved edges to the sum of conserved and non-conserved edges  (Figure \ref{fig:EC-example}) \cite{saraph2014magna}. We define a new measure of heterogeneous EC by modifying $S^3$ to account for colors of aligned end nodes of a conserved edge, as described and illustrated in Section Intro--From homogeneous to heterogeneous EC. Note that our chosen heterogeneous edge conservation weights of 1 for a fully conserved edge in which each of the two pairs of aligned nodes match in color, $\frac{2}{3}$ for a partly conserved edge in which only one of the two pairs of aligned nodes match in color, and $\frac{1}{3}$ for even less conserved edge in which none of the two pairs of aligned nodes match in color, are just one of possible choices, which we use for simplicity, as a proof-of-concept of our new heterogeneous $S^3$ measure. Other choices of weights are possible.

\subsection*{From homogeneous to heterogeneous network alignment}

We modify three recent NA methods, WAVE, MAGNA++, and SANA, to account for heterogeneous networks. We describe these algorithms and their modifications below.
\vspace{0.2cm}

\noindent\textbf{WAVE.} WAVE takes as input two networks and an NC-based matrix that captures pairwise similarities between the nodes across the compared networks, and then uses a seed-and-extend algorithm to align the networks. First, two highly similar nodes are aligned, i.e., seeded. Then, the seed's neighbors that are similar are aligned, and then the seed's neighbor's neighbors that are similar are aligned, and so on, until there is a one-to-one mapping between the networks. By aligning similar nodes, NC is optimized, and by looking at neighbors of already aligned nodes, EC is optimized, though only implicitly.

To account for heterogeneous networks, we simply plug into WAVE's alignment strategy a new matrix of node similarities that is based on our new HetNC measure generated by our proposed heterogeneous graphlet approach. Based on the fact that the algorithm looks at the neighbors of the seed, WAVE optimizes HetEC implicitly, and there is no ability to incorporate heterogeneous $S^3$ as an optimization parameter.
\vspace{0.2cm}

\noindent\textbf{MAGNA++.} MAGNA++ takes as input two networks and an NC-based matrix of node similarities, like WAVE. However, unlike WAVE, MAGNA++ uses a genetic search algorithm as its alignment strategy. MAGNA++ first starts with an initial population of randomly created alignments, the first generation. Then, high-scoring alignments (with respect to some objective function, see below) are given as input to a ``crossover" function, which combines two alignments to create a new child alignment. Many alignments from the initial population are crossed over to form new children alignments, which become the new population for the next generation. This process continues for a user-specified number of generations, and the alignment that scores the highest with respect to the objective function is given as output.

MAGNA++'s objective function can be only NC, only EC, or some combination of both. In the homogeneous case, optimizing a combination of NC (based on homogeneous graphlets) and EC ($S^3$) as objective function was shown to produce the best alignments (where the objective function is $\alpha \times \text{NC} + (1 - \alpha) \times \text{EC}$, for some $0 < \alpha < 1$; the best $\alpha$ value was determined to be 0.5)\cite{MAGNAPP}. Thus, to generalize MAGNA++ to its heterogeneous counterpart, we use MAGNA++'s alignment strategy to optimize the equally weighted combination of colored graphlet-based HetNC and heterogeneous $S^3$-based HetEC measures. To account for colored graphlet-based HetNC, we give MAGNA++ as input the colored-graphlet based node similarity matrix. To account for heterogeneous $S^3$, we modify the calculation of $S^3$ to account for node colors; source code for these changes can be found on the project website (see Abstract).

\vspace{0.2cm}

\noindent\textbf{SANA.} SANA takes as input two networks and an NC-based matrix of node similarities, like WAVE and MAGNA++, and is a search algorithm, like MAGNA++. However, it uses simulated annealing instead of a genetic algorithm as its alignment strategy. SANA starts with a single random alignment rather than a population of random alignments, and in each step it explores ``neighboring" alignments (described below). If a neighboring alignment scores higher with respect to the objective function, then it is chosen as the new alignment for the next iteration. Exploring neighboring alignments allows SANA to incrementally calculate the objective function; in particular for $S^3$, each move in the exploration process is only a small change in the alignment, and so only the changes in conserved and non-conserved edges resulting directly from the swap or change affect the $S^3$ value. Note that there is also a small chance a worse-scoring neighbor is chosen; this chance is described by the ``temperature schedule". Intuitively, the longer SANA has been running, the lower the chance of choosing a worse alignment. This continues for a set amount of time, which is a parameter of SANA. After the algorithm finishes, the alignment of the last iteration is given as output.

SANA's objective function can be only NC, only EC, or some combination of both, as is the case with MAGNA++. Thus, to generalize SANA to its heterogeneous counterpart, we use SANA's alignment strategy to optimize the equally weighted combination of colored graphlet-based HetNC and heterogeneous $S^3$-based HetEC measures. To account for colored graphlet-based HetNC, we give SANA as input the colored-graphlet based node similarity matrix. To account for heterogeneous $S^3$, we modify the incremental calculation of $S^3$ to account for node colors; pseudocode for these changes can be found on the project website (see Abstract). Note that for our heterogeneous modification of SANA we provide pseudocode rather than modified source code because SANA is not our group's method (MAGNA++ and WAVE are), and thus, there could be intellectual property restrictions regarding us sharing SANA's source code. Instead, the user can get the homogeneous SANA's code from the original authors and then modify it according to our pseudocode to allow for heterogeneous NA.

Here, we explain what a neighboring alignment means according to SANA. Let $G$ and $H$ be two networks being aligned, with $G$ having fewer nodes than $H$, and let $a, b, c, d$ be nodes in $G$, and $a', b', c', d'$ be nodes in $H$ such that $a$ is aligned to $a'$, $b$ to $b'$, $c$ to $c'$, and $d$ to $d'$. There are two kinds of neighboring alignments: swap and change. Swap neighbors differ from the original alignment in exactly two places, i.e., two source nodes in question remain the same but their images are exchanged. For example, given the existing alignment in Fig. \ref{fig:EC-example}, one of its possible swap neighbors is the alignment where $a$ is aligned to $b'$ and $b$ is aligned to $b'$, while all other aspects of the alignment remain the same. Change neighbors differ in only one place, i.e., a source node in question remains the same but its image is changed. In the example of Fig. \ref{fig:EC-example}, a possible change neighbor of the given alignment is one where $a$ is aligned to some $e'$ that initially was not part of the alignment, while all other aspects of the alignment remain the same. Consequently, if the two networks being aligned are of the same size, only swap neighbors are possible. With just these two types of neighbors, all possible alignments can potentially be reached; however, SANA focuses on those alignments that improve with respect to the objective function.

\section*{Acknowledgements}

The authors would like to thank Dr. W. Hayes for his assistance with running the homogeneous version of SANA. This work was funded by Air Force Office of Scientific Research (AFOSR) Young Investigator Research Program (YIP) under award number FA9550-16-1-0147, and National Science Foundation (NSF) Faculty Early Career Development Program (CAREER) under award number CCF-1452795.

\section*{Author contributions statement}

SG, FEF, and TM designed the study. JJ, FEF, and TM developed the proposed heterogeneous graphlet approach. SG, FEF, and TM developed the proposed HetNA approaches, including the proposed HetNC and HetEC measures. SG, JJ, and FEF implemented the proposed approaches; the GUI part of the implementation was carried out solely by JJ. SG and FEF performed the computational experiments and produced all results. SG, FEF, and TM analyzed the results. All authors wrote, read, and approved the paper.

\section*{Competing interests}

The authors declare that they have no competing interests.

\bibliographystyle{abbrv}

\newpage

\newcommand{\beginsupplement}{%
        \setcounter{table}{0}
        \renewcommand{\thetable}{S\arabic{table}}%
        \setcounter{figure}{0}
        \renewcommand{\thefigure}{S\arabic{figure}}%
        \renewcommand{\figurename}{Supplementary Figure}

     }

\beginsupplement

\section*{Supplementary information for From homogeneous to heterogeneous network alignment via colored graphlets}


\begin{figure}[h]
    \subfloat[]{\includegraphics[width=0.30\textwidth]{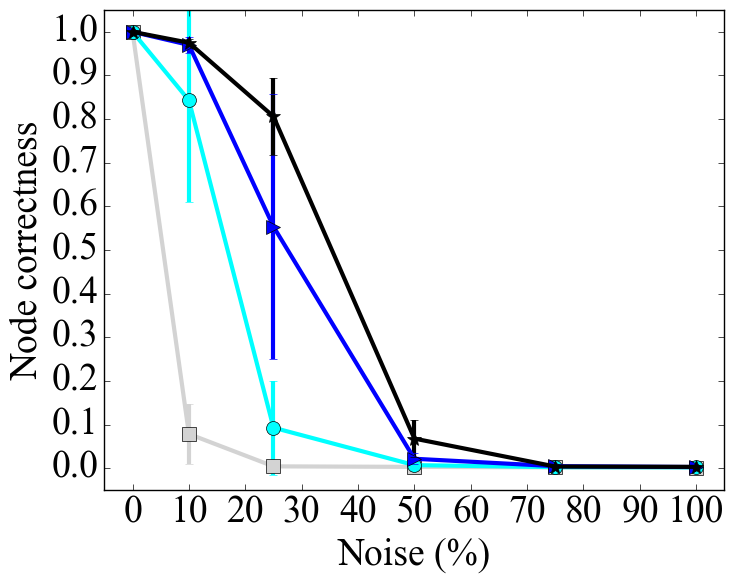}}
    \subfloat[]{\includegraphics[width=0.28\textwidth]{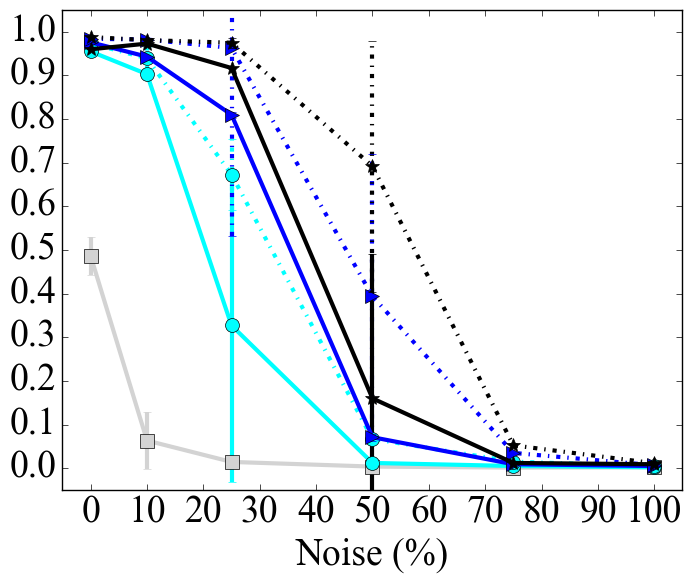}}
    \subfloat[]{\includegraphics[width=0.43\textwidth]{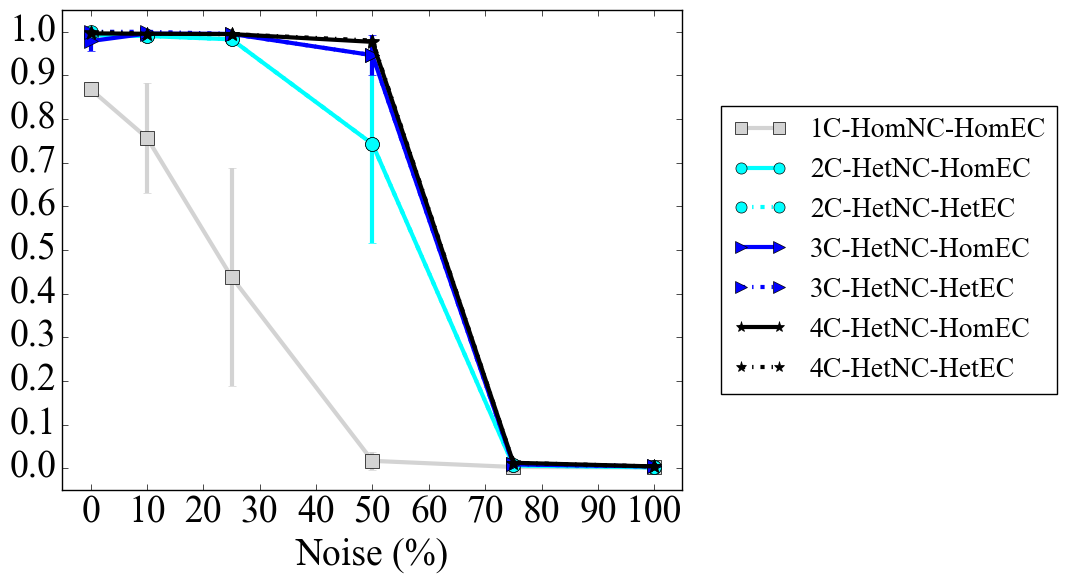}}
    
    \caption{\label{fig:supp-geo}Detailed alignment quality results regarding the effect of the \textbf{number of node colors} on alignment quality as a function of noise level for \textbf{synthetic, specifically geometric}, networks using (a) WAVE, (b) MAGNA++, and (c) SANA. Gray squares, light blue circles, dark blue triangles, and black stars indicate the aligned networks containing one, two, three, and four node colors, respectively. For two or more node colors, solid lines represent using HetNC-HomEC, and dashed lines represent using HetNC-HetEC.}
    
\end{figure}

\begin{figure}[h]
    \subfloat[]{\includegraphics[width=0.30\textwidth]{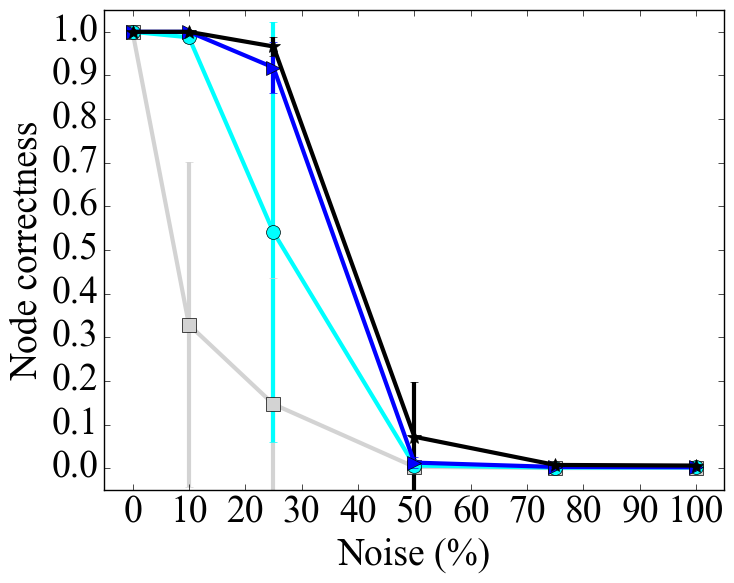}}
    \subfloat[]{\includegraphics[width=0.28\textwidth]{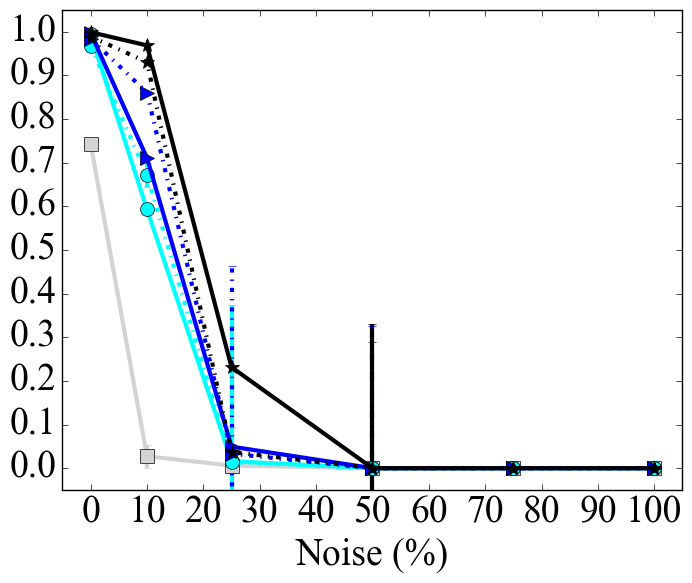}}
    \subfloat[]{\includegraphics[width=0.43\textwidth]{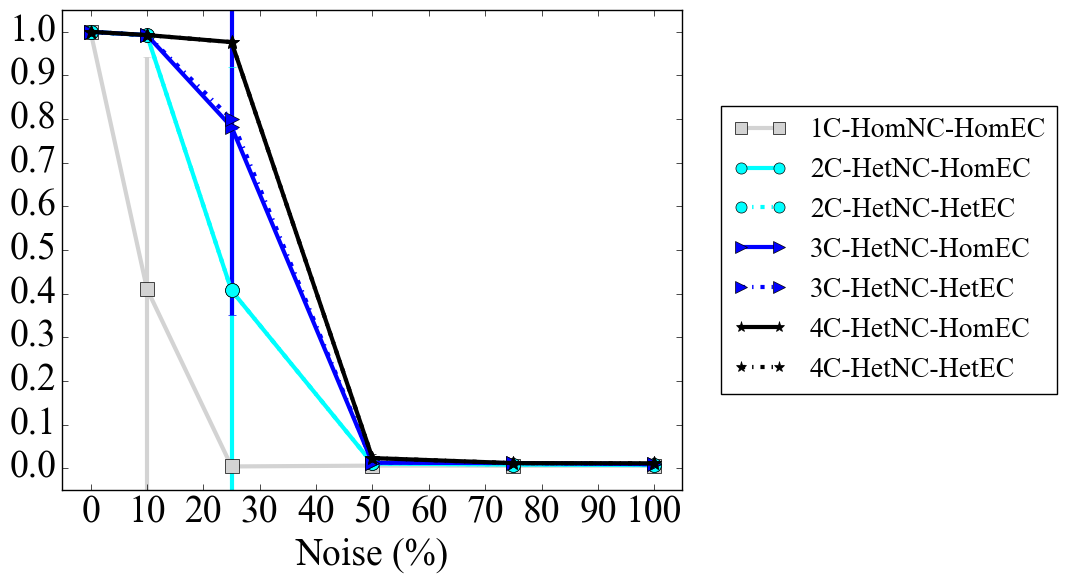}}
    
    \caption{\label{fig:supp-sf}Detailed alignment quality results regarding the effect of the \textbf{number of node colors} on alignment quality as a function of noise level for \textbf{synthetic, specifically scale-free}, networks using (a) WAVE, (b) MAGNA++, and (c) SANA. The figure can be interpreted in the same way as Supplementary Figure \ref{fig:supp-geo}.}
    
\end{figure}

\begin{figure}[h]
    \subfloat[]{\includegraphics[width=0.33\textwidth]{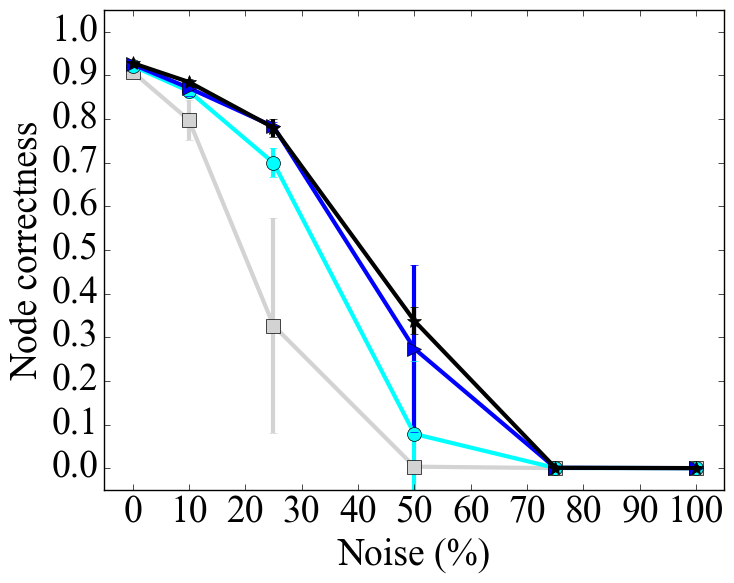}}
    \subfloat[]{\includegraphics[width=0.48\textwidth]{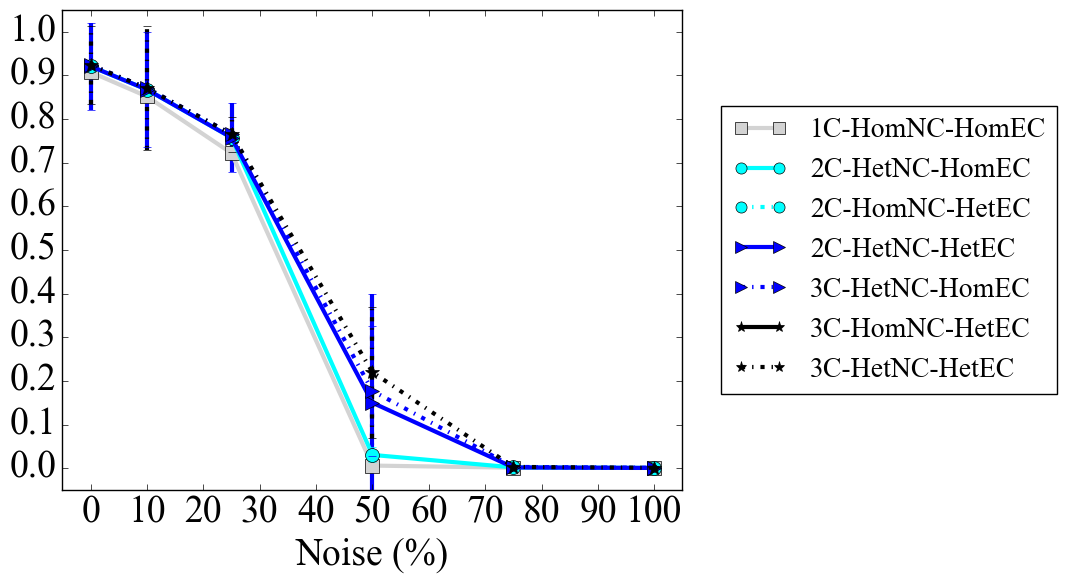}}

    \caption{\label{fig:supp-apmsexpr}Detailed alignment quality results regarding the effect of the \textbf{number of node colors} on alignment quality as a function of noise level for \textbf{PPI, specifically APMS-Expr}, networks using (a) WAVE and (b) SANA. The figure can be interpreted in the same way as Supplementary Figure \ref{fig:supp-geo}. Recall that for these larger networks, we have not run MAGNA++ due to its high computational complexity.}
    
\end{figure}

\begin{figure}[h]
    \subfloat[]{\includegraphics[width=0.33\textwidth]{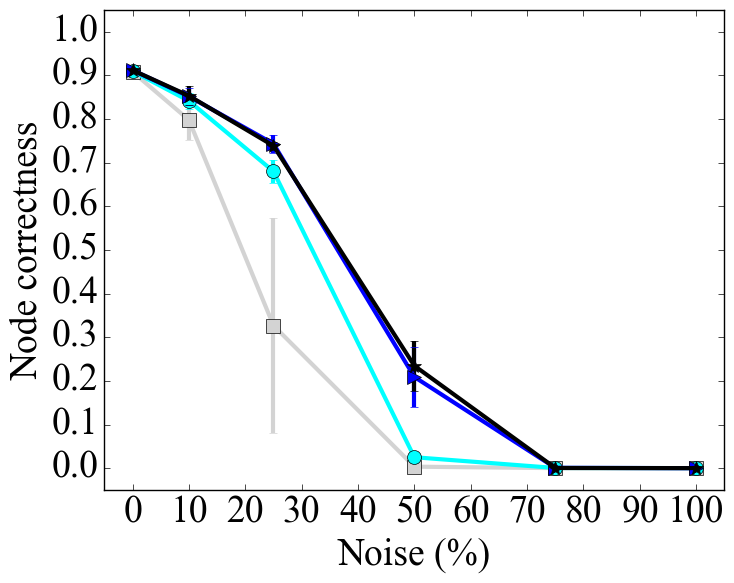}}
    \subfloat[]{\includegraphics[width=0.48\textwidth]{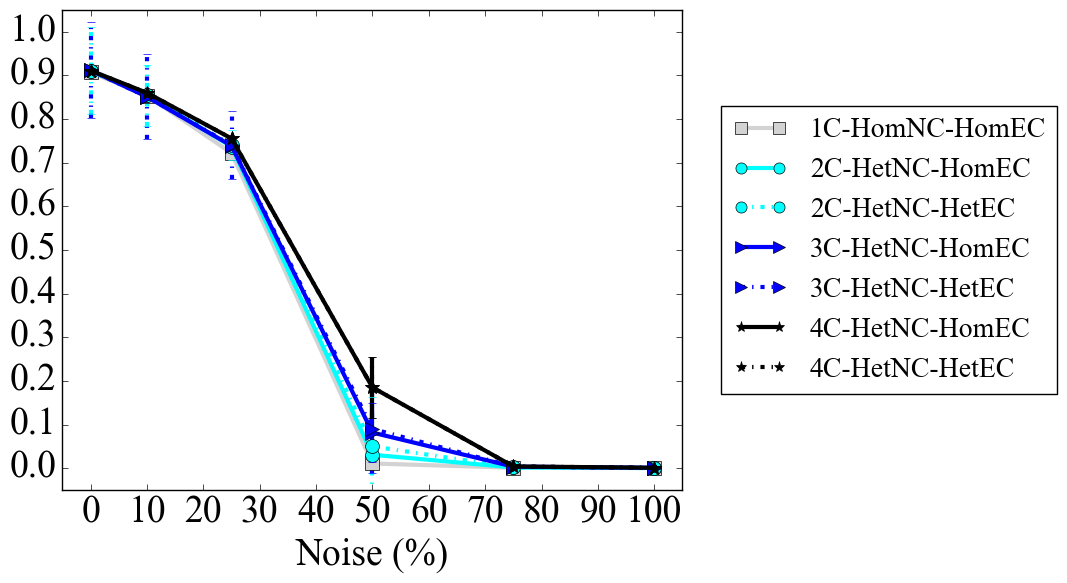}}

    \caption{\label{fig:supp-apmsseq}Detailed alignment quality results regarding the effect of the \textbf{number of node colors} on alignment quality as a function of noise level for \textbf{PPI, specifically APMS-Seq}, networks using (a) WAVE and (b) SANA. The figure can be interpreted in the same way as Supplementary Figure \ref{fig:supp-geo}. Recall that for these larger networks, we have not run MAGNA++ due to its high computational complexity.}
    
\end{figure}

\begin{figure}[h]
    \subfloat[]{\includegraphics[width=0.33\textwidth]{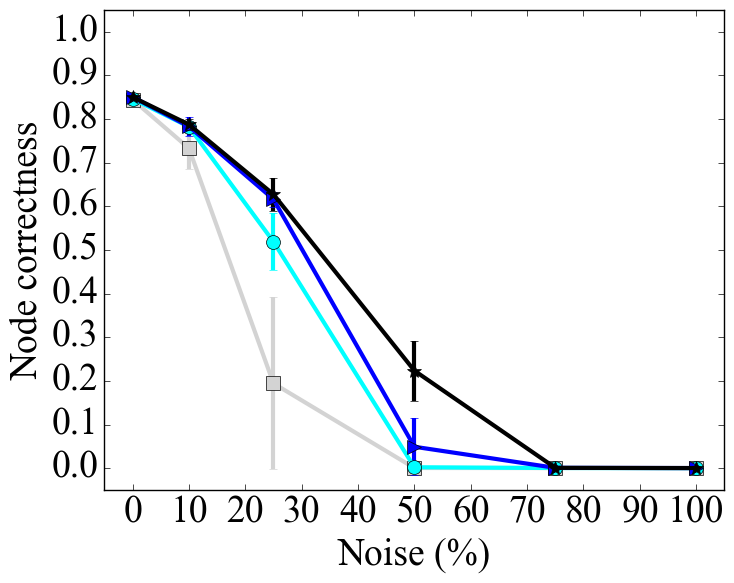}}
    \subfloat[]{\includegraphics[width=0.48\textwidth]{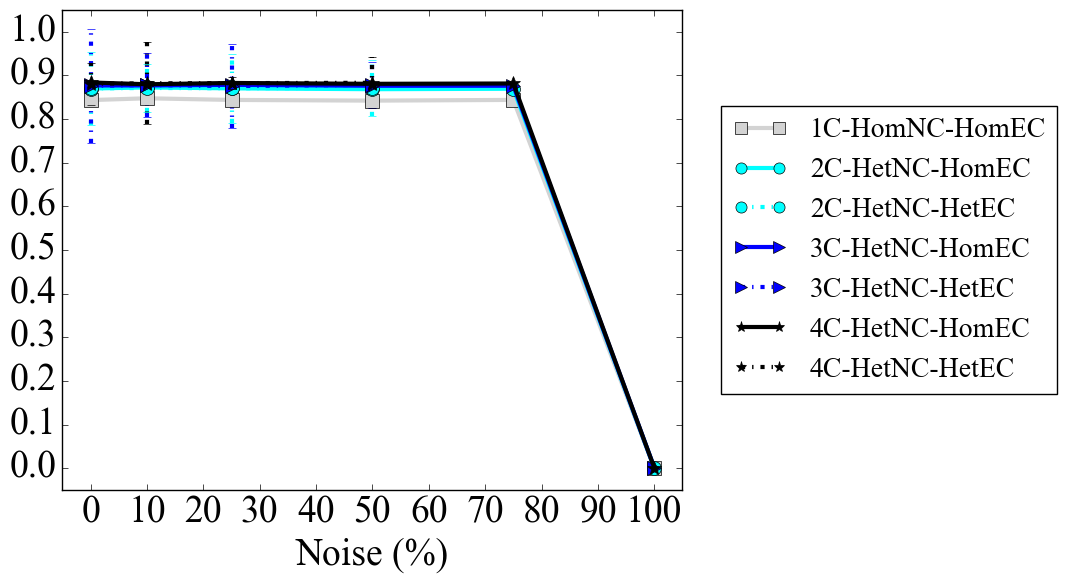}}

    \caption{\label{fig:supp-y2hexpr}Detailed alignment quality results regarding the effect of the \textbf{number of node colors} on alignment quality as a function of noise level for \textbf{PPI, specifically Y2H-Expr}, networks using (a) WAVE and (b) SANA. The figure can be interpreted in the same way as Supplementary Figure \ref{fig:supp-geo}. Recall that for these larger networks, we have not run MAGNA++ due to its high computational complexity.}
    
\end{figure}

\begin{figure}[h]
    \subfloat[]{\includegraphics[width=0.33\textwidth]{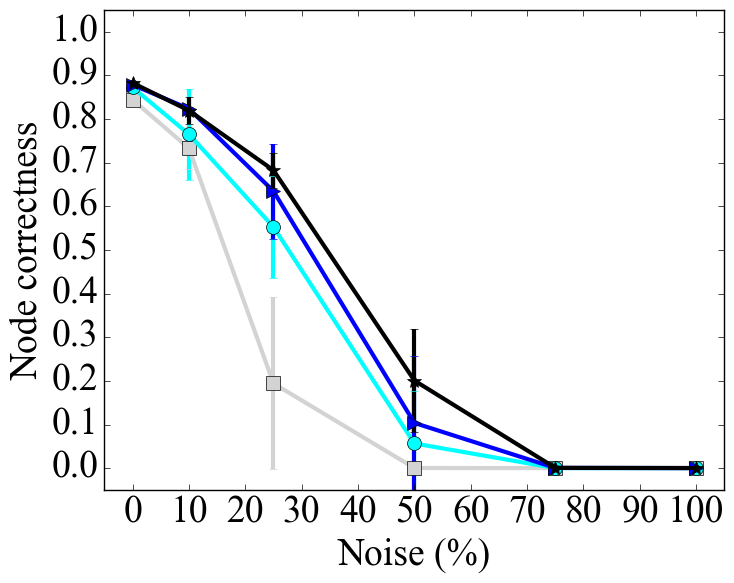}}
    \subfloat[]{\includegraphics[width=0.48\textwidth]{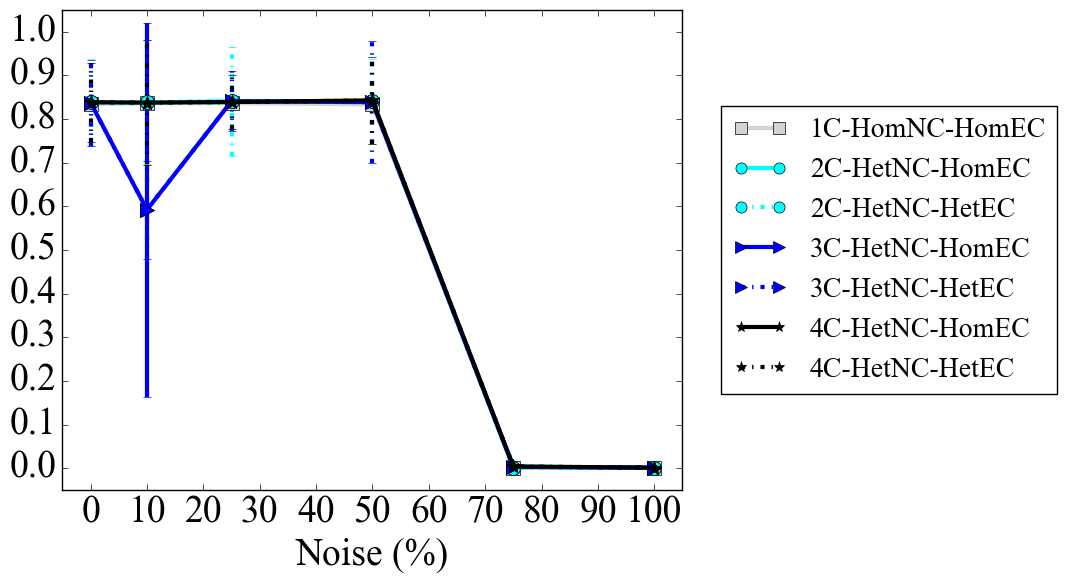}}

    \caption{\label{fig:supp-y2hseq}Detailed alignment quality results regarding the effect of the \textbf{number of node colors} on alignment quality as a function of noise level for \textbf{PPI, specifically Y2H-Seq}, networks using (a) WAVE and (b) SANA. The figure can be interpreted in the same way as Supplementary Figure \ref{fig:supp-geo}. Recall that for these larger networks, we have not run MAGNA++ due to its high computational complexity.}
    
\end{figure}

\begin{figure}[h]
    \subfloat[]{\includegraphics[width=0.33\textwidth]{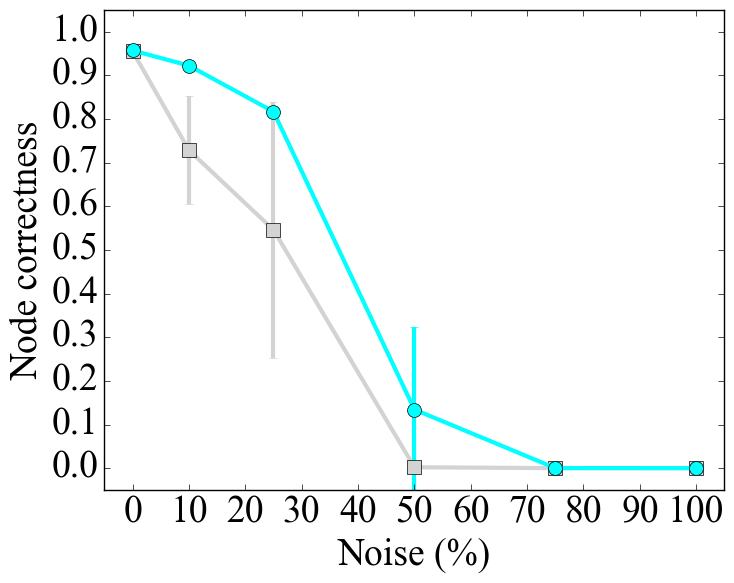}}
    \subfloat[]{\includegraphics[width=0.48\textwidth]{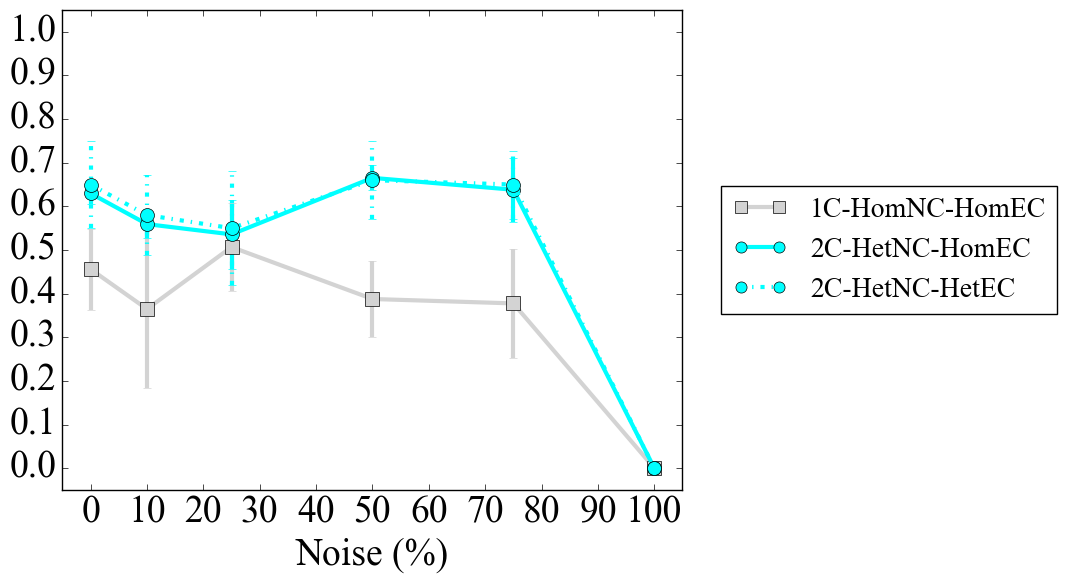}}

    \caption{\label{fig:supp-pg-apms}Detailed alignment quality results regarding the effect of the \textbf{number of node colors} on alignment quality as a function of noise level for \textbf{protein-GO, specifically protein-GO-APMS}, networks using (a) WAVE and (b) SANA. The figure can be interpreted in the same way as Supplementary Figure \ref{fig:supp-geo}. Recall that for these larger networks, we have not run MAGNA++ due to its high computational complexity.}
    
\end{figure}

\begin{figure}[h]
    \subfloat[]{\includegraphics[width=0.33\textwidth]{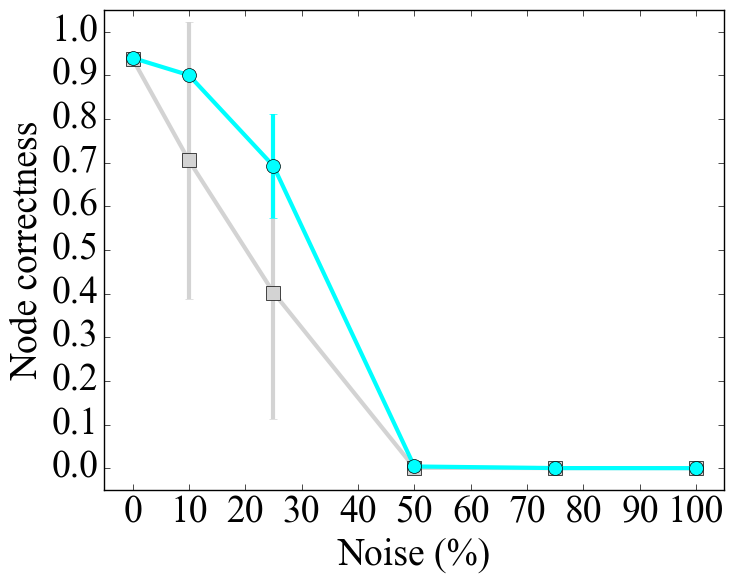}}
    \subfloat[]{\includegraphics[width=0.48\textwidth]{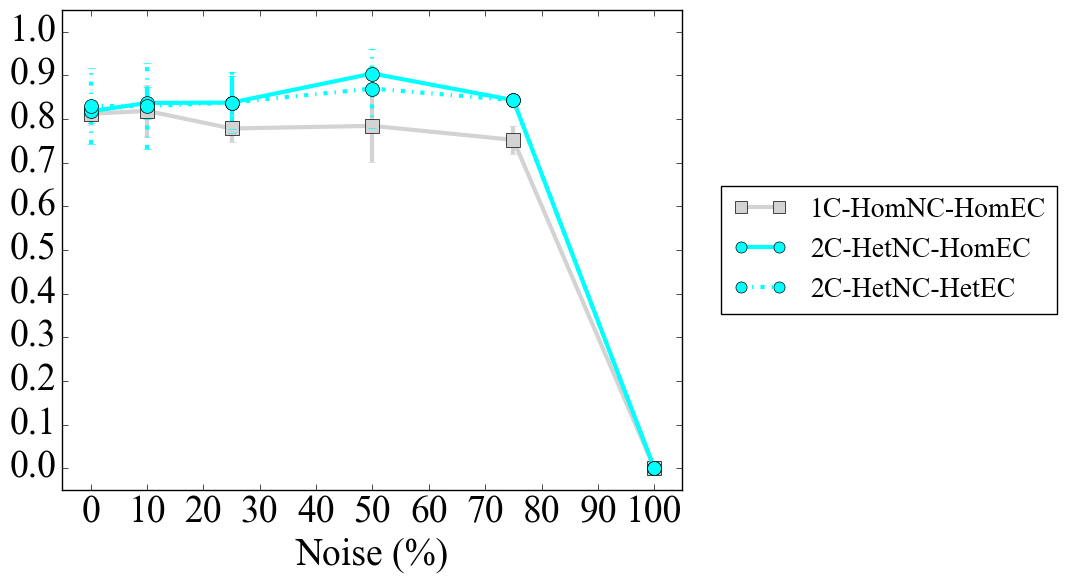}}

    \caption{\label{fig:supp-pg-y2h}Detailed alignment quality results regarding the effect of the \textbf{number of node colors} on alignment quality as a function of noise level for \textbf{protein-GO, specifically protein-GO-Y2H}, networks using (a) WAVE and (b) SANA. The figure can be interpreted in the same way as Supplementary Figure \ref{fig:supp-geo}. Recall that for these larger networks, we have not run MAGNA++ due to its high computational complexity.}
    
\end{figure}

\begin{figure}[h]
    \centering
    \subfloat[]{\includegraphics[width=0.345\textwidth]{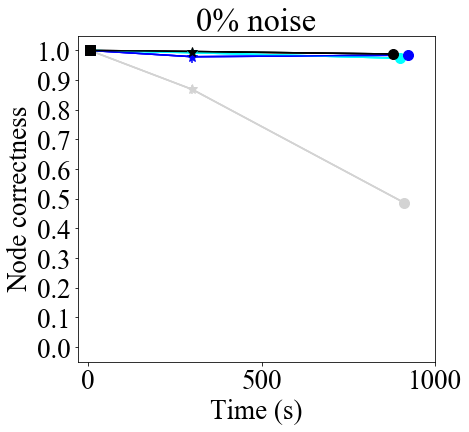}}
    \subfloat[]{\includegraphics[width=0.325\textwidth]{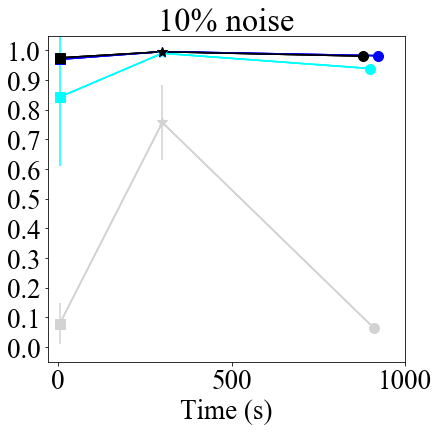}}\newline 
    \subfloat[]{\includegraphics[width=0.345\textwidth]{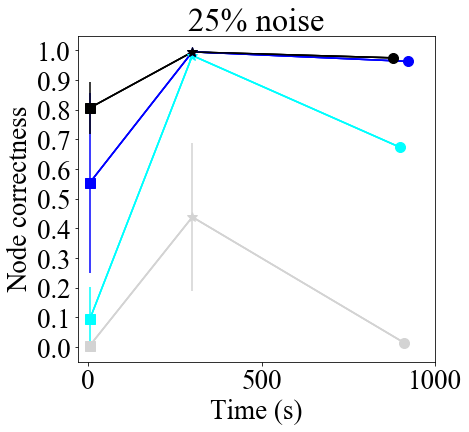}} \subfloat[]{\includegraphics[width=0.325\textwidth]{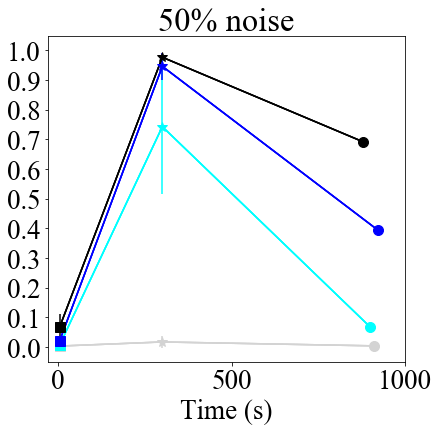}}\newline
    \subfloat[]{\includegraphics[width=0.345\textwidth]{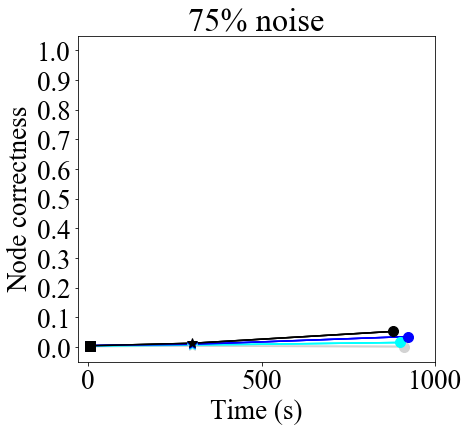}}
    \subfloat[]{\includegraphics[width=0.435\textwidth]{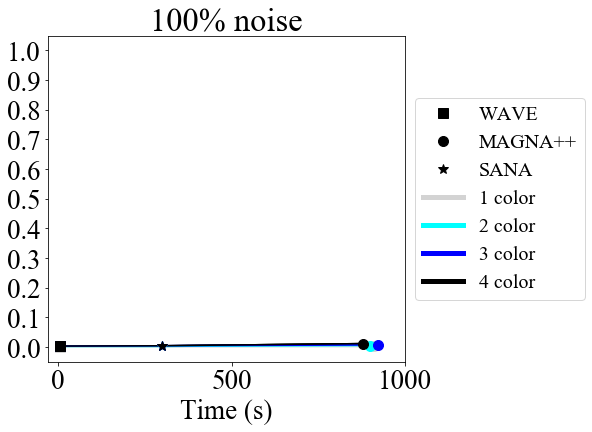}}
    \caption{\label{fig:supp-time-geo}Detailed results comparing the \textbf{running time} and effect of the \textbf{number of node colors} for different methods for all tested noise levels on \textbf{synthetic, specifically geometric}, networks. The \textit{x}-axis the the running time of the method, and the \textit{y}-axis is the alignment quality. Here we use different shapes to represent the different methods and different colored lines to represent how many node colors are used. Lines are drawn between methods using the same number of colors.}
\end{figure}

\begin{figure}[h]
    \centering
    \subfloat[]{\includegraphics[width=0.345\textwidth]{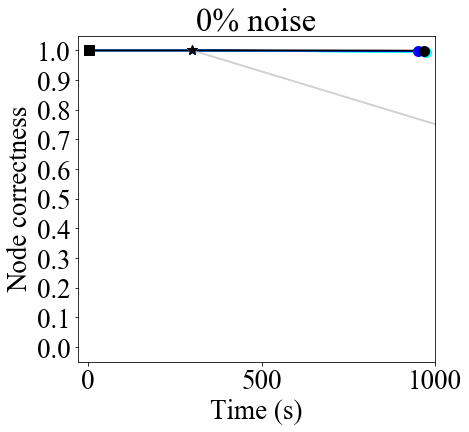}}
    \subfloat[]{\includegraphics[width=0.325\textwidth]{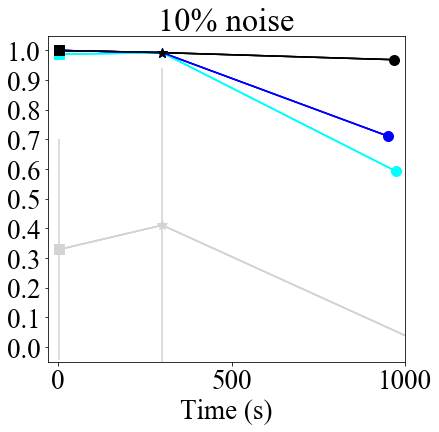}}\newline 
    \subfloat[]{\includegraphics[width=0.345\textwidth]{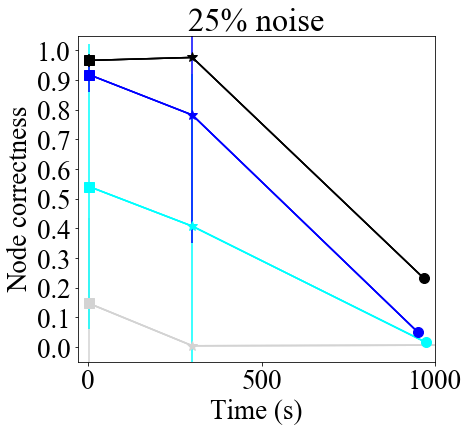}} \subfloat[]{\includegraphics[width=0.325\textwidth]{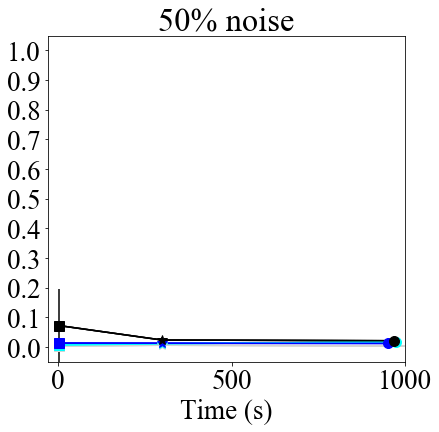}}\newline
    \subfloat[]{\includegraphics[width=0.345\textwidth]{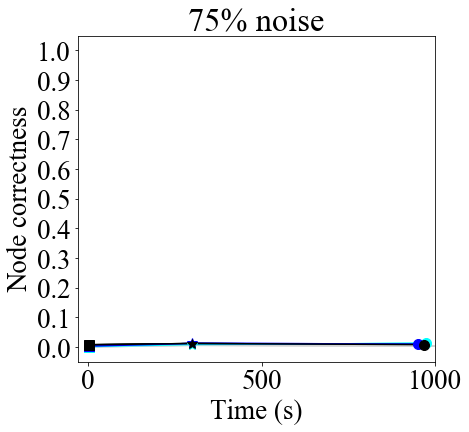}}
    \subfloat[]{\includegraphics[width=0.435\textwidth]{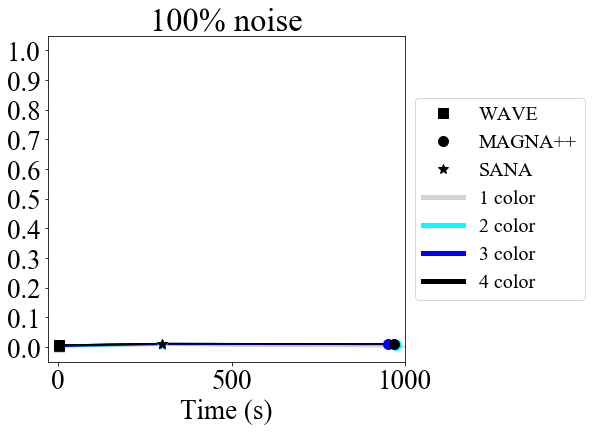}}
    \caption{\label{fig:supp-time-sf}Detailed results comparing the \textbf{running time} and effect of the \textbf{number of node colors} for different methods for all tested noise levels on \textbf{synthetic, specifically scale-free}, networks. The figure can be interpreted in the same way as Supplementary Figure \ref{fig:supp-time-geo}.}
\end{figure}

\begin{figure}[h]
    \centering
    \subfloat[]{\includegraphics[width=0.345\textwidth]{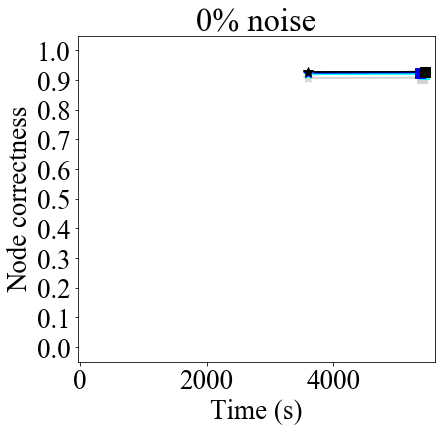}}
    \subfloat[]{\includegraphics[width=0.325\textwidth]{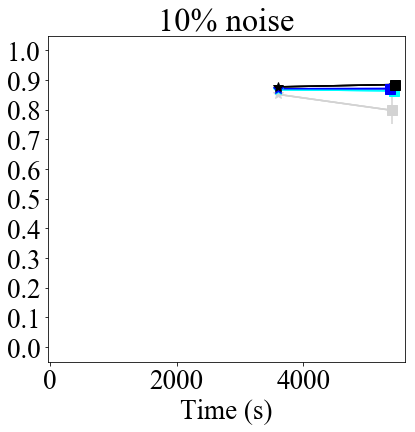}}\newline 
    \subfloat[]{\includegraphics[width=0.345\textwidth]{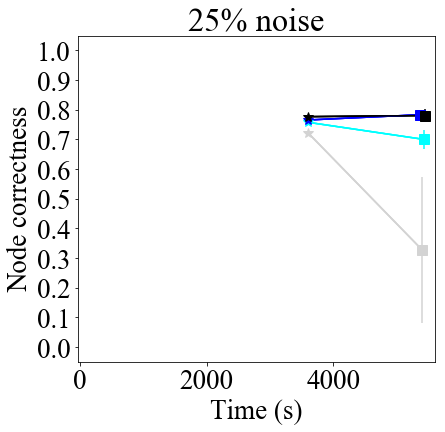}} \subfloat[]{\includegraphics[width=0.325\textwidth]{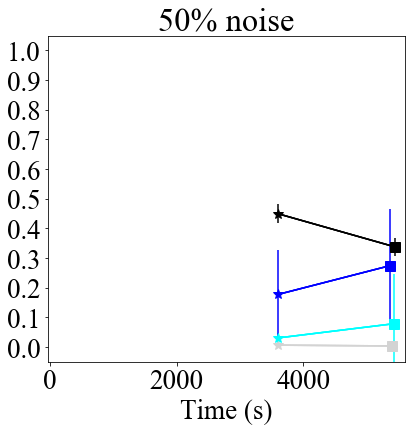}}\newline
    \subfloat[]{\includegraphics[width=0.345\textwidth]{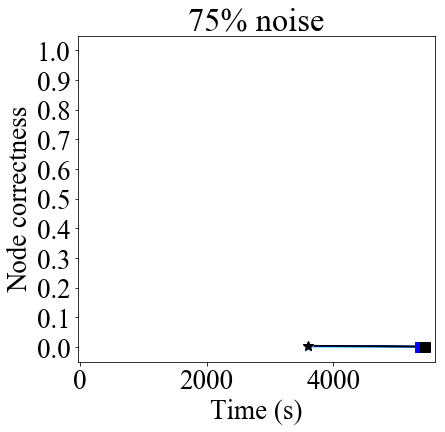}}
    \subfloat[]{\includegraphics[width=0.435\textwidth]{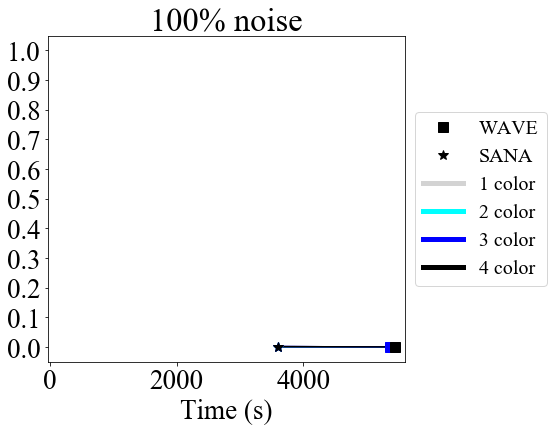}}
    \caption{\label{fig:supp-time-apms-expr}Detailed results comparing the \textbf{running time} and effect of the \textbf{number of node colors} for different methods for all tested noise levels on \textbf{PPI, specifically APMS-Expr}, networks. The figure can be interpreted in the same way as Supplementary Figure \ref{fig:supp-time-geo}.}
\end{figure}

\begin{figure}[h]
    \centering
    \subfloat[]{\includegraphics[width=0.345\textwidth]{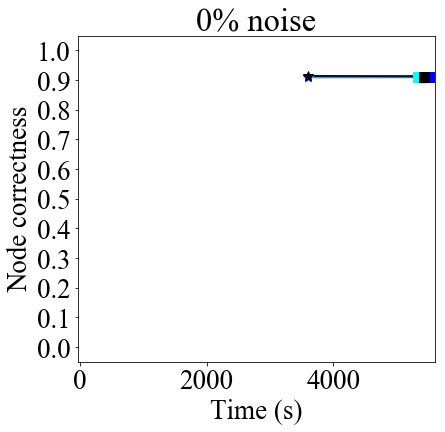}}
    \subfloat[]{\includegraphics[width=0.325\textwidth]{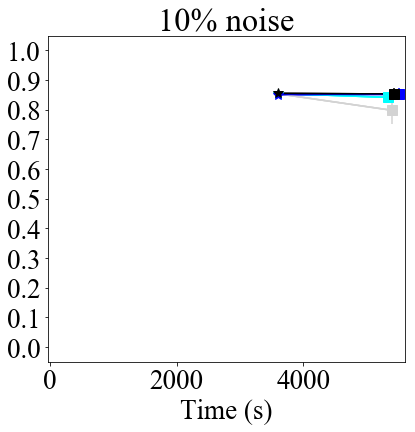}}\newline 
    \subfloat[]{\includegraphics[width=0.345\textwidth]{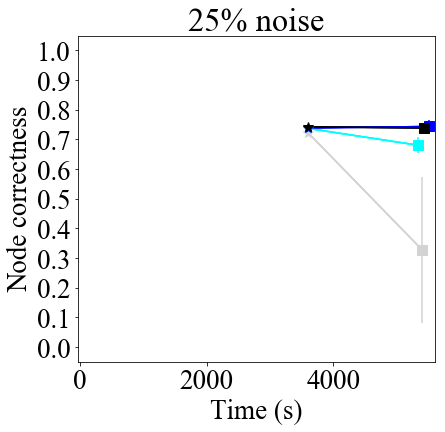}} \subfloat[]{\includegraphics[width=0.325\textwidth]{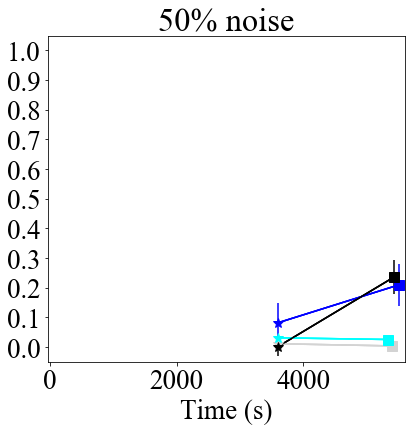}}\newline
    \subfloat[]{\includegraphics[width=0.345\textwidth]{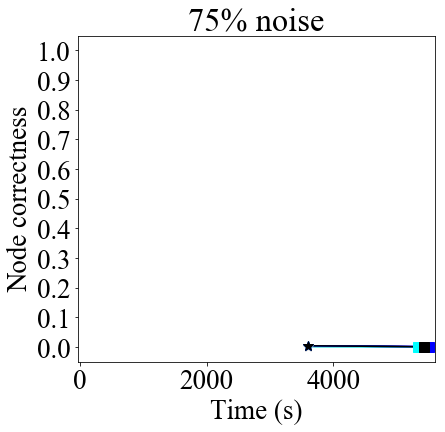}}
    \subfloat[]{\includegraphics[width=0.435\textwidth]{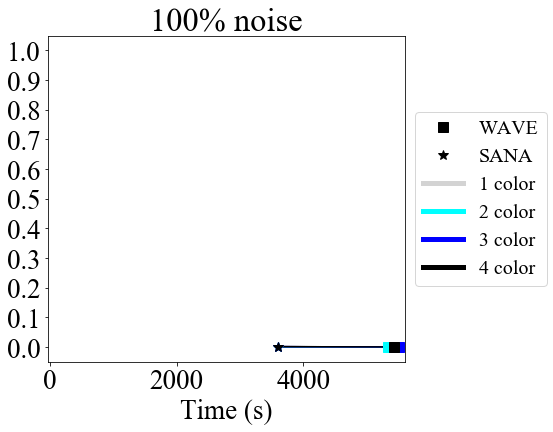}}
    \caption{\label{fig:supp-time-apms-seq}Detailed results comparing the \textbf{running time} and effect of the \textbf{number of node colors} for different methods for all tested noise levels on \textbf{PPI, specifically APMS-Seq}, networks. The figure can be interpreted in the same way as Supplementary Figure \ref{fig:supp-time-geo}.}
\end{figure}

\begin{figure}[h]
    \centering
    \subfloat[]{\includegraphics[width=0.345\textwidth]{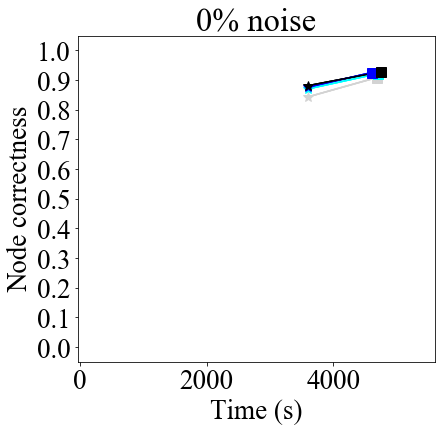}}
    \subfloat[]{\includegraphics[width=0.325\textwidth]{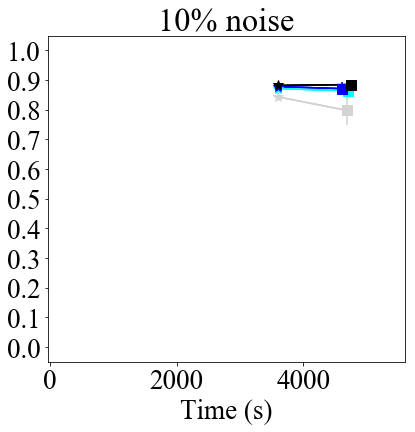}}\newline 
    \subfloat[]{\includegraphics[width=0.345\textwidth]{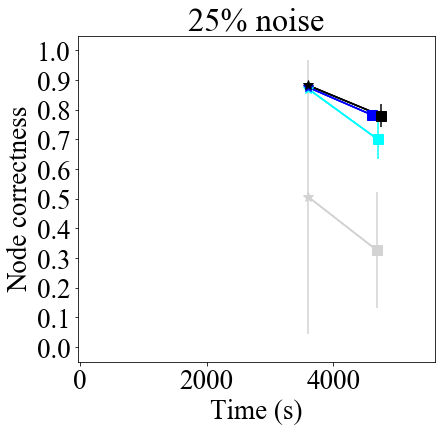}} \subfloat[]{\includegraphics[width=0.325\textwidth]{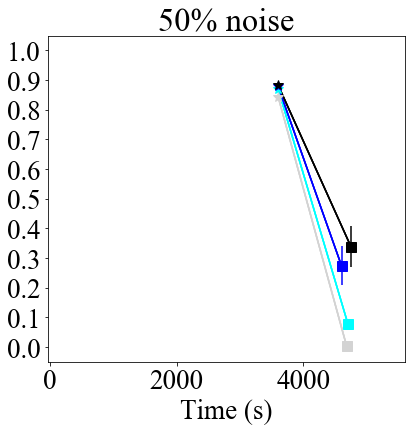}}\newline
    \subfloat[]{\includegraphics[width=0.345\textwidth]{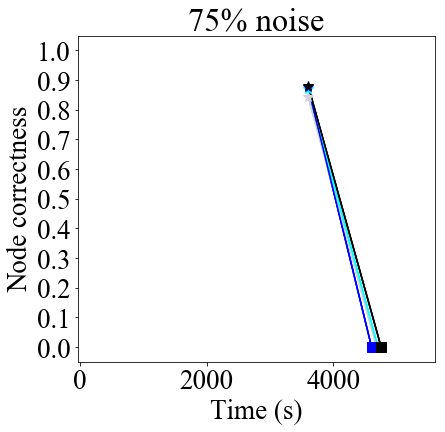}}
    \subfloat[]{\includegraphics[width=0.435\textwidth]{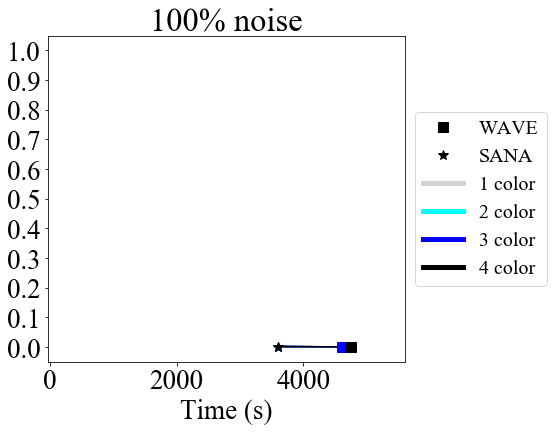}}
    \caption{\label{fig:supp-time-y2h-expr}Detailed results comparing the \textbf{running time} and effect of the \textbf{number of node colors} for different methods for all tested noise levels on \textbf{PPI, specifically Y2H-Expr}, networks. The figure can be interpreted in the same way as Supplementary Figure \ref{fig:supp-time-geo}.}
\end{figure}

\begin{figure}[h]
    \centering
    \subfloat[]{\includegraphics[width=0.345\textwidth]{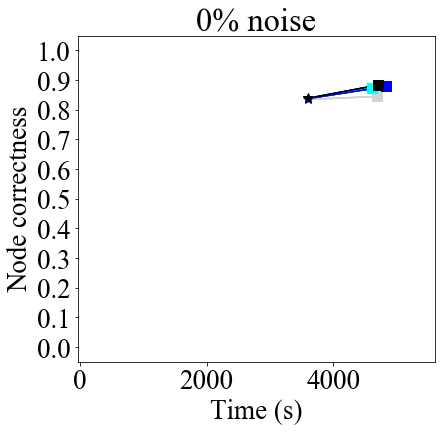}}
    \subfloat[]{\includegraphics[width=0.325\textwidth]{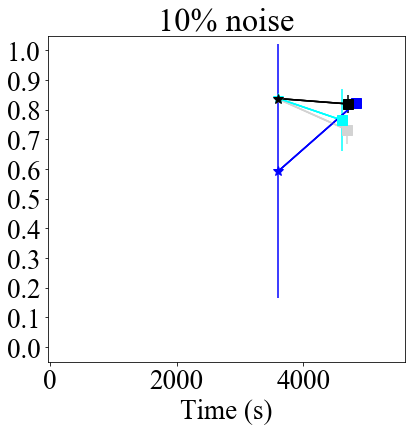}}\newline 
    \subfloat[]{\includegraphics[width=0.345\textwidth]{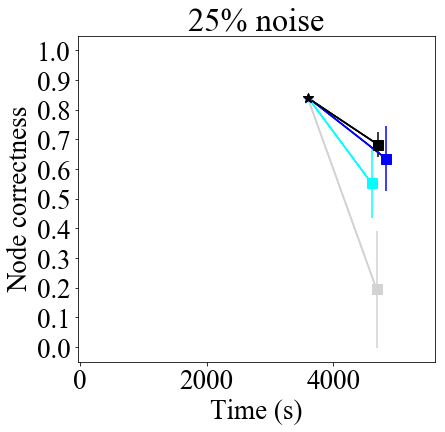}} \subfloat[]{\includegraphics[width=0.325\textwidth]{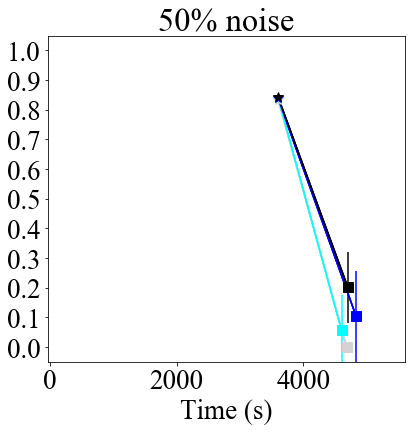}}\newline
    \subfloat[]{\includegraphics[width=0.345\textwidth]{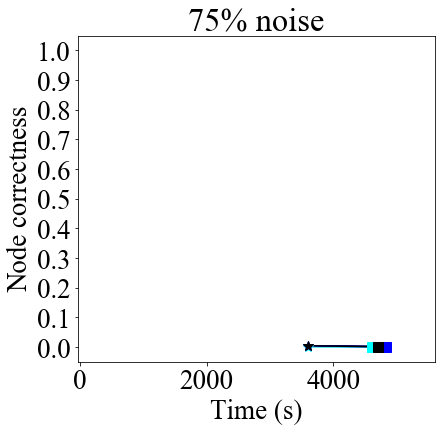}}
    \subfloat[]{\includegraphics[width=0.435\textwidth]{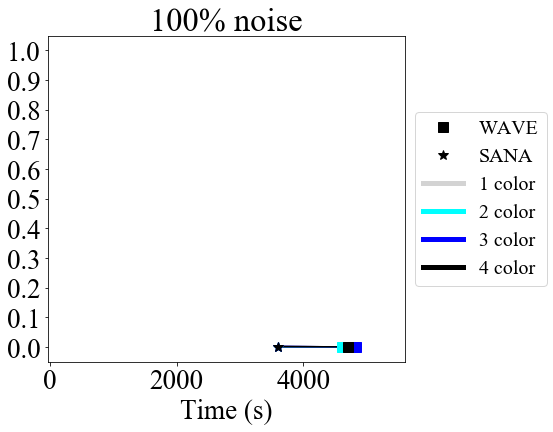}}
    \caption{\label{fig:supp-time-y2h-seq}Detailed results comparing the \textbf{running time} and effect of the \textbf{number of node colors} for different methods for all tested noise levels on \textbf{PPI, specifically Y2H-Seq}, networks. The figure can be interpreted in the same way as Supplementary Figure \ref{fig:supp-time-geo}.}
\end{figure}

\begin{figure}[h]
    \centering
    \subfloat[]{\includegraphics[width=0.345\textwidth]{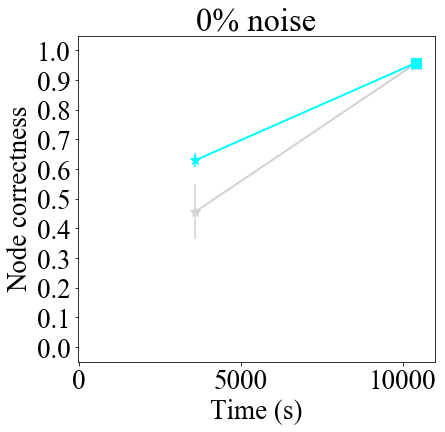}}
    \subfloat[]{\includegraphics[width=0.325\textwidth]{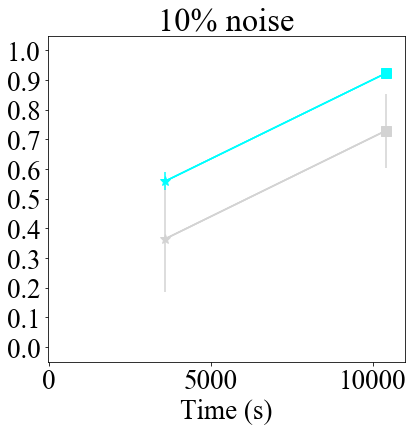}}\newline 
    \subfloat[]{\includegraphics[width=0.345\textwidth]{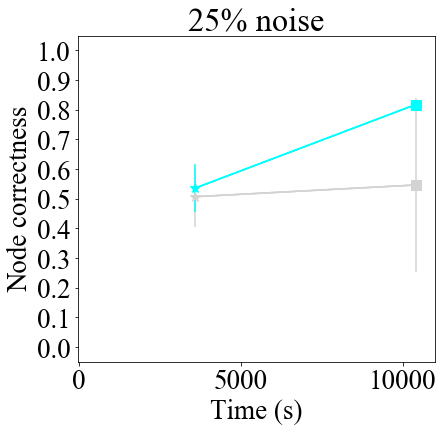}} \subfloat[]{\includegraphics[width=0.325\textwidth]{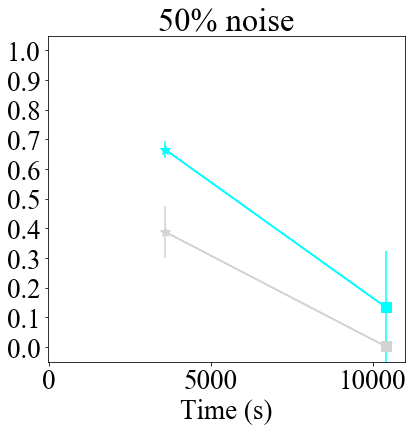}}\newline
    \subfloat[]{\includegraphics[width=0.345\textwidth]{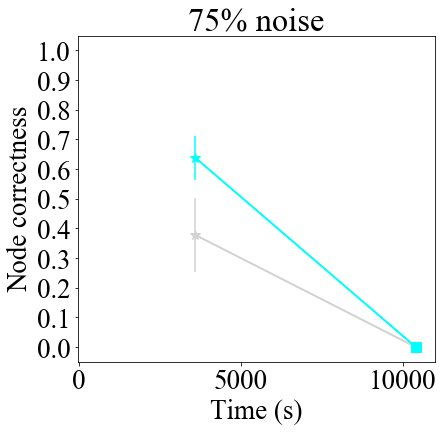}}
    \subfloat[]{\includegraphics[width=0.435\textwidth]{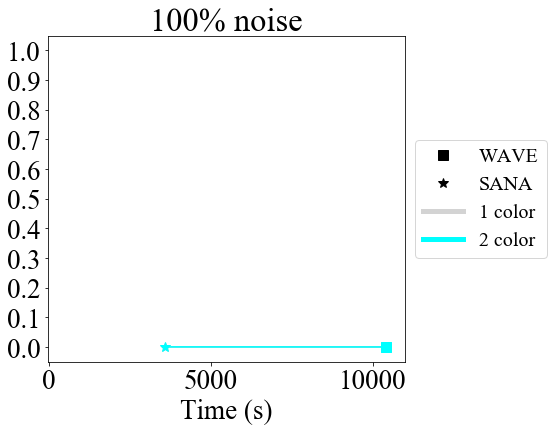}}
    \caption{\label{fig:supp-time-pg-apms}Detailed results comparing the \textbf{running time} and effect of the \textbf{number of node colors} for different methods for all tested noise levels on \textbf{protein-GO, specifically protein-GO-APMS}, networks. The figure can be interpreted in the same way as Supplementary Figure \ref{fig:supp-time-geo}.}
\end{figure}

\begin{figure}[h]
    \centering
    \subfloat[]{\includegraphics[width=0.345\textwidth]{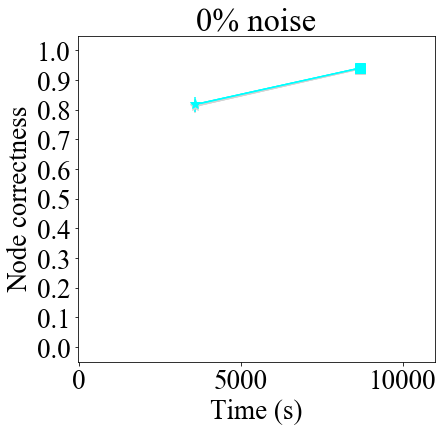}}
    \subfloat[]{\includegraphics[width=0.325\textwidth]{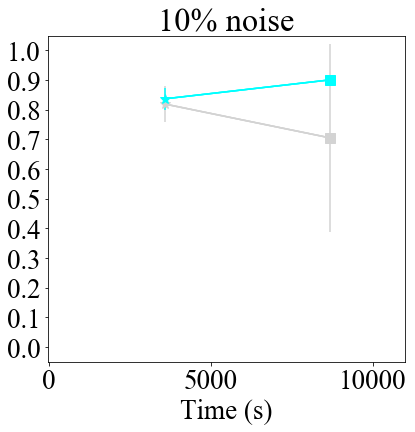}}\newline 
    \subfloat[]{\includegraphics[width=0.345\textwidth]{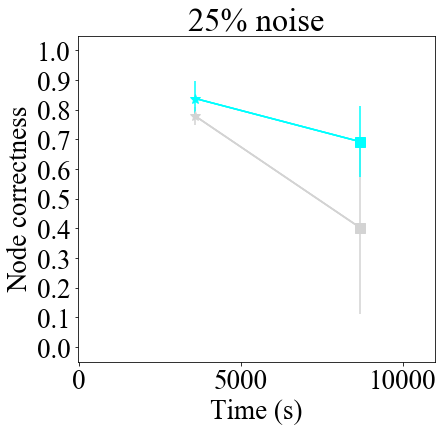}} \subfloat[]{\includegraphics[width=0.325\textwidth]{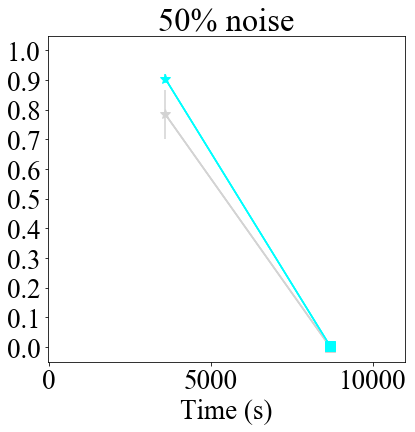}}\newline
    \subfloat[]{\includegraphics[width=0.345\textwidth]{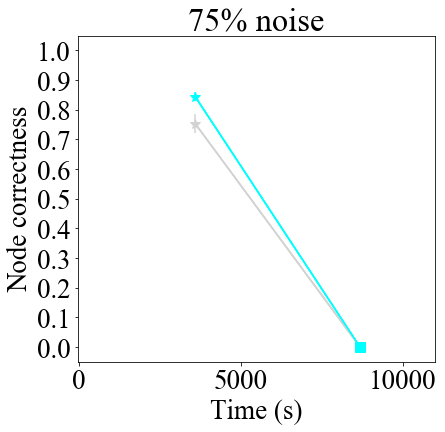}}
    \subfloat[]{\includegraphics[width=0.435\textwidth]{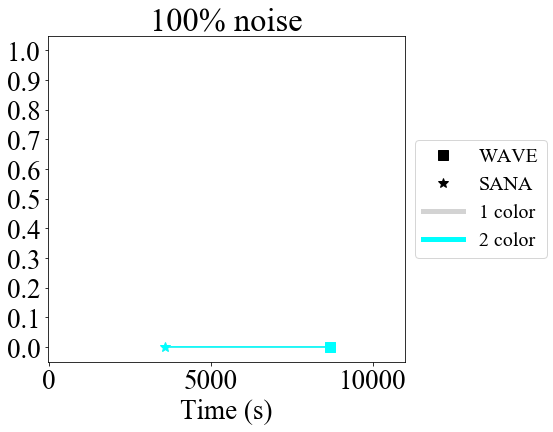}}
    \caption{\label{fig:supp-time-pg-y2h}Detailed results comparing the \textbf{running time} and effect of the \textbf{number of node colors} for different methods for all tested noise levels on \textbf{protein-GO, specifically protein-GO-Y2H}, networks. The figure can be interpreted in the same way as Supplementary Figure \ref{fig:supp-time-geo}.}
\end{figure}

\end{document}


\newcommand{\beginsupplement}{%
        \setcounter{table}{0}
        \renewcommand{\thetable}{S\arabic{table}}%
        \setcounter{figure}{0}
        \renewcommand{\thefigure}{S\arabic{figure}}%
        \renewcommand{\figurename}{Supplementary Figure}

     }

\beginsupplement

\section*{Supplementary information for From homogeneous to heterogeneous network alignment via colored graphlets}
\subsection*{Shawn Gu$^1$, John Johnson$^1$, Fazle E. Faisal$^{1, 2}$, and Tijana Milenkovi\'{c}$^{1, 2, *}$}

$^1$ Department of Computer Science and Engineering, University of Notre Dame, Notre Dame, IN, 46556, USA

\noindent$^2$ Eck Institute for Global Health and Interdisciplinary Center for Network Science and Applications (iCeNSA), University of Notre Dame, Notre Dame, IN, 46556, USA

\noindent$^*$ To whom correspondence should be addressed (email: tmilenko@nd.edu)

\author[1]{Shawn Gu}
\author[1]{John Johnson}
\author[1,2]{Fazle E. Faisal}
\author[1,2,*]{Tijana Milenkovi\'{c}}
\affil[1]{Department of Computer Science and Engineering, University of Notre Dame, Notre Dame, IN, 46556, USA}
\affil[2]{Eck Institute for Global Health and Interdisciplinary Center for Network Science and Applications (iCeNSA), University of Notre Dame, Notre Dame, IN, 46556, USA}

\affil[*]{To whom correspondence should be addressed (email: tmilenko@nd.edu)}

\vspace{10mm}
\noindent Here, we include all results.


\begin{figure}[h]
    \subfloat[]{\includegraphics[width=0.30\textwidth]{img/supp/wavegeo.png}}
    \subfloat[]{\includegraphics[width=0.28\textwidth]{img/supp/magnappgeo.png}}
    \subfloat[]{\includegraphics[width=0.43\textwidth]{img/supp/sanageo.png}}
    
    \caption{\label{fig:supp-geo}Detailed alignment quality results regarding the effect of the \textbf{number of node colors} on alignment quality as a function of noise level for \textbf{synthetic, specifically geometric}, networks using (a) WAVE, (b) MAGNA++, and (c) SANA. Gray squares, light blue circles, dark blue triangles, and black stars indicate the aligned networks containing one, two, three, and four node colors, respectively. For two or more node colors, solid lines represent using HetNC-HomEC, and dashed lines represent using HetNC-HetEC.}
    
\end{figure}

\begin{figure}[h]
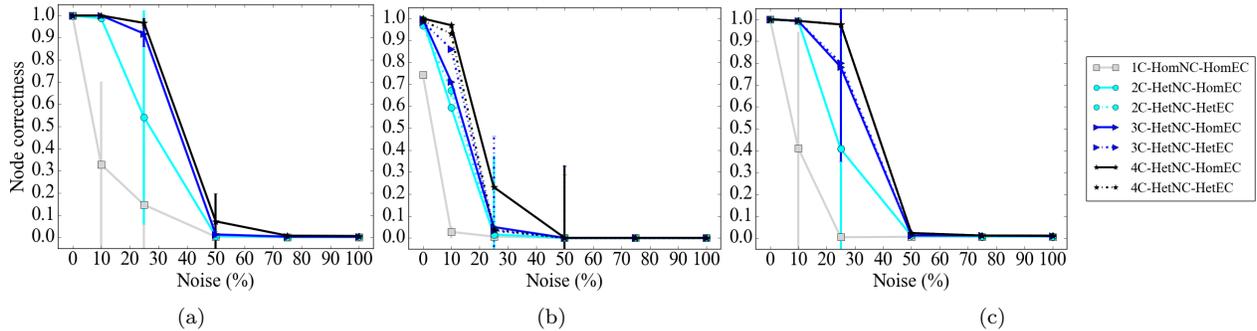

    \subfloat[]{\includegraphics[width=0.30\textwidth]{img/supp/wavesf.png}}
    \subfloat[]{\includegraphics[width=0.28\textwidth]{img/supp/magnappsf.png}}
    \subfloat[]{\includegraphics[width=0.43\textwidth]{img/supp/sanasf.png}}
    
    \caption{\label{fig:supp-sf}Detailed alignment quality results regarding the effect of the \textbf{number of node colors} on alignment quality as a function of noise level for \textbf{synthetic, specifically scale-free}, networks using (a) WAVE, (b) MAGNA++, and (c) SANA. The figure can be interpreted in the same way as Supplementary Figure \ref{fig:supp-geo}.}
    
\end{figure}

\begin{figure}[h]
    \subfloat[]{\includegraphics[width=0.33\textwidth]{img/supp/waveapmsexpr.png}}
    \subfloat[]{\includegraphics[width=0.48\textwidth]{img/supp/sanaapmsexpr.png}}

    \caption{\label{fig:supp-apmsexpr}Detailed alignment quality results regarding the effect of the \textbf{number of node colors} on alignment quality as a function of noise level for \textbf{PPI, specifically APMS-Expr}, networks using (a) WAVE and (b) SANA. The figure can be interpreted in the same way as Supplementary Figure \ref{fig:supp-geo}. Recall that for these larger networks, we have not run MAGNA++ due to its high computational complexity.}
    
\end{figure}

\begin{figure}[h]
    \subfloat[]{\includegraphics[width=0.33\textwidth]{img/supp/waveapmsseq.png}}
    \subfloat[]{\includegraphics[width=0.48\textwidth]{img/supp/sanaapmsseq.png}}

    \caption{\label{fig:supp-apmsseq}Detailed alignment quality results regarding the effect of the \textbf{number of node colors} on alignment quality as a function of noise level for \textbf{PPI, specifically APMS-Seq}, networks using (a) WAVE and (b) SANA. The figure can be interpreted in the same way as Supplementary Figure \ref{fig:supp-geo}. Recall that for these larger networks, we have not run MAGNA++ due to its high computational complexity.}
    
\end{figure}

\begin{figure}[h]
    \subfloat[]{\includegraphics[width=0.33\textwidth]{img/supp/wavey2hexpr.png}}
    \subfloat[]{\includegraphics[width=0.48\textwidth]{img/supp/sanay2hexpr.png}}

    \caption{\label{fig:supp-y2hexpr}Detailed alignment quality results regarding the effect of the \textbf{number of node colors} on alignment quality as a function of noise level for \textbf{PPI, specifically Y2H-Expr}, networks using (a) WAVE and (b) SANA. The figure can be interpreted in the same way as Supplementary Figure \ref{fig:supp-geo}. Recall that for these larger networks, we have not run MAGNA++ due to its high computational complexity.}
    
\end{figure}

\begin{figure}[h]
    \subfloat[]{\includegraphics[width=0.33\textwidth]{img/supp/wavey2hseq.png}}
    \subfloat[]{\includegraphics[width=0.48\textwidth]{img/supp/sanay2hseq.png}}

    \caption{\label{fig:supp-y2hseq}Detailed alignment quality results regarding the effect of the \textbf{number of node colors} on alignment quality as a function of noise level for \textbf{PPI, specifically Y2H-Seq}, networks using (a) WAVE and (b) SANA. The figure can be interpreted in the same way as Supplementary Figure \ref{fig:supp-geo}. Recall that for these larger networks, we have not run MAGNA++ due to its high computational complexity.}
    
\end{figure}

\begin{figure}[h]
    \subfloat[]{\includegraphics[width=0.33\textwidth]{img/supp/wave-pg-apms.png}}
    \subfloat[]{\includegraphics[width=0.48\textwidth]{img/supp/sana-pg-apms.png}}

    \caption{\label{fig:supp-pg-apms}Detailed alignment quality results regarding the effect of the \textbf{number of node colors} on alignment quality as a function of noise level for \textbf{protein-GO, specifically protein-GO-APMS}, networks using (a) WAVE and (b) SANA. The figure can be interpreted in the same way as Supplementary Figure \ref{fig:supp-geo}. Recall that for these larger networks, we have not run MAGNA++ due to its high computational complexity.}
    
\end{figure}

\begin{figure}[h]
    \subfloat[]{\includegraphics[width=0.33\textwidth]{img/supp/wave-pg-y2h.png}}
    \subfloat[]{\includegraphics[width=0.48\textwidth]{img/supp/sana-pg-y2h.png}}

    \caption{\label{fig:supp-pg-y2h}Detailed alignment quality results regarding the effect of the \textbf{number of node colors} on alignment quality as a function of noise level for \textbf{protein-GO, specifically protein-GO-Y2H}, networks using (a) WAVE and (b) SANA. The figure can be interpreted in the same way as Supplementary Figure \ref{fig:supp-geo}. Recall that for these larger networks, we have not run MAGNA++ due to its high computational complexity.}
    
\end{figure}

\begin{figure}[h]
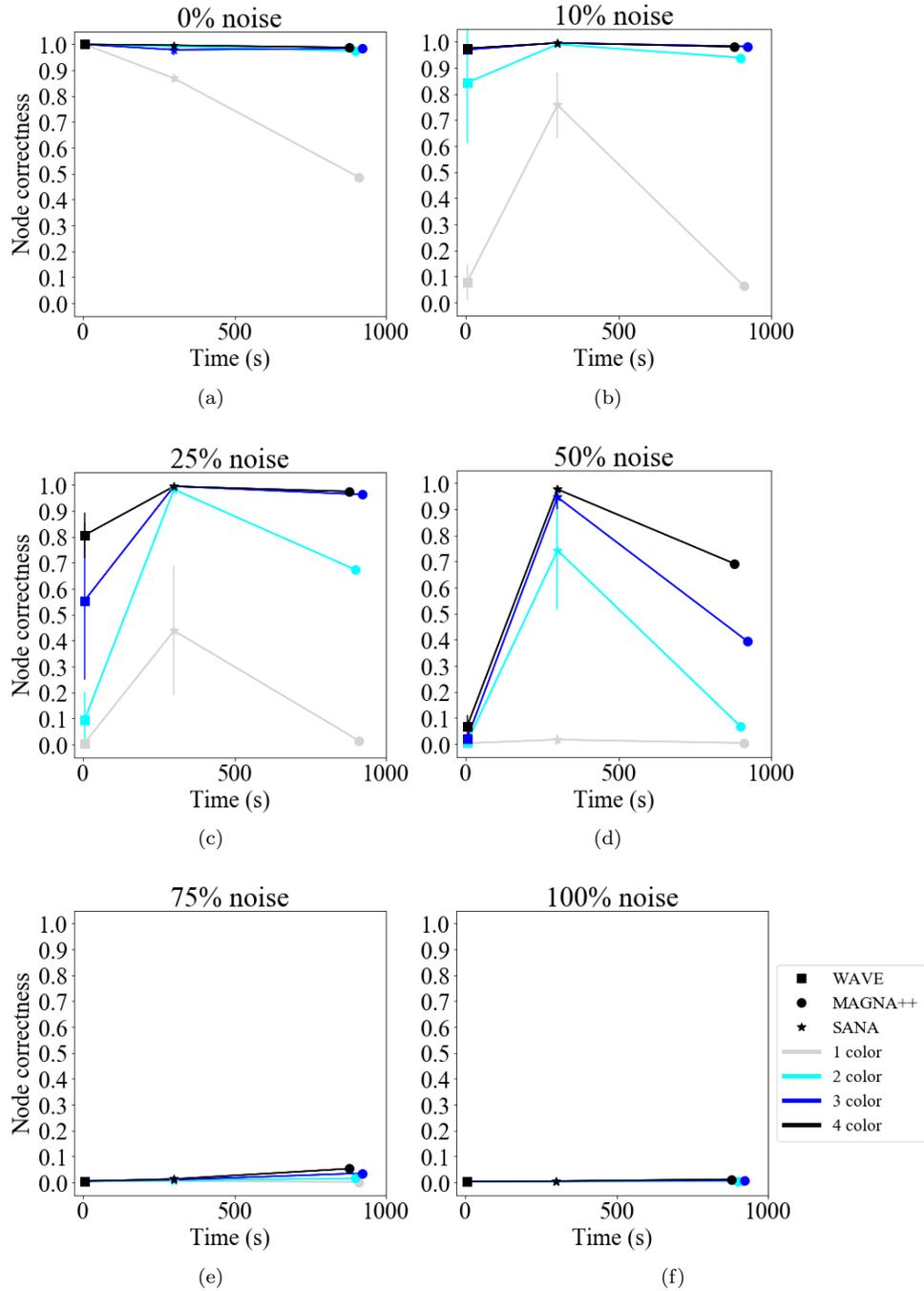

    \centering
    \subfloat[]{\includegraphics[width=0.345\textwidth]{img/supp/geo/time-geo0.png}}
    \subfloat[]{\includegraphics[width=0.325\textwidth]{img/supp/geo/time-geo10.png}}\newline 
    \subfloat[]{\includegraphics[width=0.345\textwidth]{img/supp/geo/time-geo25.png}} \subfloat[]{\includegraphics[width=0.325\textwidth]{img/supp/geo/time-geo50.png}}\newline
    \subfloat[]{\includegraphics[width=0.345\textwidth]{img/supp/geo/time-geo75.png}}
    \subfloat[]{\includegraphics[width=0.435\textwidth]{img/supp/geo/time-geo100.png}}
    \caption{\label{fig:supp-time-geo}Detailed results comparing the \textbf{running time} and effect of the \textbf{number of node colors} for different methods for all tested noise levels on \textbf{synthetic, specifically geometric}, networks. The \textit{x}-axis the the running time of the method, and the \textit{y}-axis is the alignment quality. Here we use different shapes to represent the different methods and different colored lines to represent how many node colors are used. Lines are drawn between methods using the same number of colors.}
\end{figure}

\begin{figure}[h]
    \centering
    \subfloat[]{\includegraphics[width=0.345\textwidth]{img/supp/sf/time-sf0.png}}
    \subfloat[]{\includegraphics[width=0.325\textwidth]{img/supp/sf/time-sf10.png}}\newline 
    \subfloat[]{\includegraphics[width=0.345\textwidth]{img/supp/sf/time-sf25.png}} \subfloat[]{\includegraphics[width=0.325\textwidth]{img/supp/sf/time-sf50.png}}\newline
    \subfloat[]{\includegraphics[width=0.345\textwidth]{img/supp/sf/time-sf75.png}}
    \subfloat[]{\includegraphics[width=0.435\textwidth]{img/supp/sf/time-sf100.png}}
    \caption{\label{fig:supp-time-sf}Detailed results comparing the \textbf{running time} and effect of the \textbf{number of node colors} for different methods for all tested noise levels on \textbf{synthetic, specifically scale-free}, networks. The figure can be interpreted in the same way as Supplementary Figure \ref{fig:supp-time-geo}.}
\end{figure}

\begin{figure}[h]
    \centering
    \subfloat[]{\includegraphics[width=0.345\textwidth]{img/supp/apms-expr/time0.png}}
    \subfloat[]{\includegraphics[width=0.325\textwidth]{img/supp/apms-expr/time10.png}}\newline 
    \subfloat[]{\includegraphics[width=0.345\textwidth]{img/supp/apms-expr/time25.png}} \subfloat[]{\includegraphics[width=0.325\textwidth]{img/supp/apms-expr/time50.png}}\newline
    \subfloat[]{\includegraphics[width=0.345\textwidth]{img/supp/apms-expr/time75.png}}
    \subfloat[]{\includegraphics[width=0.435\textwidth]{img/supp/apms-expr/time100.png}}
    \caption{\label{fig:supp-time-apms-expr}Detailed results comparing the \textbf{running time} and effect of the \textbf{number of node colors} for different methods for all tested noise levels on \textbf{PPI, specifically APMS-Expr}, networks. The figure can be interpreted in the same way as Supplementary Figure \ref{fig:supp-time-geo}.}
\end{figure}

\begin{figure}[h]
    \centering
    \subfloat[]{\includegraphics[width=0.345\textwidth]{img/supp/apms-seq/time0.png}}
    \subfloat[]{\includegraphics[width=0.325\textwidth]{img/supp/apms-seq/time10.png}}\newline 
    \subfloat[]{\includegraphics[width=0.345\textwidth]{img/supp/apms-seq/time25.png}} \subfloat[]{\includegraphics[width=0.325\textwidth]{img/supp/apms-seq/time50.png}}\newline
    \subfloat[]{\includegraphics[width=0.345\textwidth]{img/supp/apms-seq/time75.png}}
    \subfloat[]{\includegraphics[width=0.435\textwidth]{img/supp/apms-seq/time100.png}}
    \caption{\label{fig:supp-time-apms-seq}Detailed results comparing the \textbf{running time} and effect of the \textbf{number of node colors} for different methods for all tested noise levels on \textbf{PPI, specifically APMS-Seq}, networks. The figure can be interpreted in the same way as Supplementary Figure \ref{fig:supp-time-geo}.}
\end{figure}

\begin{figure}[h]
    \centering
    \subfloat[]{\includegraphics[width=0.345\textwidth]{img/supp/y2h-expr/time0.png}}
    \subfloat[]{\includegraphics[width=0.325\textwidth]{img/supp/y2h-expr/time10.png}}\newline 
    \subfloat[]{\includegraphics[width=0.345\textwidth]{img/supp/y2h-expr/time25.png}} \subfloat[]{\includegraphics[width=0.325\textwidth]{img/supp/y2h-expr/time50.png}}\newline
    \subfloat[]{\includegraphics[width=0.345\textwidth]{img/supp/y2h-expr/time75.png}}
    \subfloat[]{\includegraphics[width=0.435\textwidth]{img/supp/y2h-expr/time100.png}}
    \caption{\label{fig:supp-time-y2h-expr}Detailed results comparing the \textbf{running time} and effect of the \textbf{number of node colors} for different methods for all tested noise levels on \textbf{PPI, specifically Y2H-Expr}, networks. The figure can be interpreted in the same way as Supplementary Figure \ref{fig:supp-time-geo}.}
\end{figure}

\begin{figure}[h]
    \centering
    \subfloat[]{\includegraphics[width=0.345\textwidth]{img/supp/y2h-seq/time0.png}}
    \subfloat[]{\includegraphics[width=0.325\textwidth]{img/supp/y2h-seq/time10.png}}\newline 
    \subfloat[]{\includegraphics[width=0.345\textwidth]{img/supp/y2h-seq/time25.png}} \subfloat[]{\includegraphics[width=0.325\textwidth]{img/supp/y2h-seq/time50.png}}\newline
    \subfloat[]{\includegraphics[width=0.345\textwidth]{img/supp/y2h-seq/time75.png}}
    \subfloat[]{\includegraphics[width=0.435\textwidth]{img/supp/y2h-seq/time100.png}}
    \caption{\label{fig:supp-time-y2h-seq}Detailed results comparing the \textbf{running time} and effect of the \textbf{number of node colors} for different methods for all tested noise levels on \textbf{PPI, specifically Y2H-Seq}, networks. The figure can be interpreted in the same way as Supplementary Figure \ref{fig:supp-time-geo}.}
\end{figure}

\begin{figure}[h]
    \centering
    \subfloat[]{\includegraphics[width=0.345\textwidth]{img/supp/pg-apms/time0.png}}
    \subfloat[]{\includegraphics[width=0.325\textwidth]{img/supp/pg-apms/time10.png}}\newline 
    \subfloat[]{\includegraphics[width=0.345\textwidth]{img/supp/pg-apms/time25.png}} \subfloat[]{\includegraphics[width=0.325\textwidth]{img/supp/pg-apms/time50.png}}\newline
    \subfloat[]{\includegraphics[width=0.345\textwidth]{img/supp/pg-apms/time75.png}}
    \subfloat[]{\includegraphics[width=0.435\textwidth]{img/supp/pg-apms/time100.png}}
    \caption{\label{fig:supp-time-pg-apms}Detailed results comparing the \textbf{running time} and effect of the \textbf{number of node colors} for different methods for all tested noise levels on \textbf{protein-GO, specifically protein-GO-APMS}, networks. The figure can be interpreted in the same way as Supplementary Figure \ref{fig:supp-time-geo}.}
\end{figure}

\begin{figure}[h]
    \centering
    \subfloat[]{\includegraphics[width=0.345\textwidth]{img/supp/pg-y2h/time0.png}}
    \subfloat[]{\includegraphics[width=0.325\textwidth]{img/supp/pg-y2h/time10.png}}\newline 
    \subfloat[]{\includegraphics[width=0.345\textwidth]{img/supp/pg-y2h/time25.png}} \subfloat[]{\includegraphics[width=0.325\textwidth]{img/supp/pg-y2h/time50.png}}\newline
    \subfloat[]{\includegraphics[width=0.345\textwidth]{img/supp/pg-y2h/time75.png}}
    \subfloat[]{\includegraphics[width=0.435\textwidth]{img/supp/pg-y2h/time100.png}}
    \caption{\label{fig:supp-time-pg-y2h}Detailed results comparing the \textbf{running time} and effect of the \textbf{number of node colors} for different methods for all tested noise levels on \textbf{protein-GO, specifically protein-GO-Y2H}, networks. The figure can be interpreted in the same way as Supplementary Figure \ref{fig:supp-time-geo}.}
\end{figure}